\documentclass{aa}  

\usepackage{graphicx}
\usepackage{txfonts}
\usepackage{gensymb}
\usepackage{booktabs}
\usepackage{array}
\usepackage{pifont}

\newcommand{\figref}[2][\figurename~]{#1\ref{#2}}
\newcommand{\tabref}[2][\tablename~]{#1\ref{#2}}

\newcommand{\secref}[2][Section~]{#1\ref{#2}}

\begin{document}

   \title{Implications of different stellar spectra for the climate of tidally--locked Earth-like exoplanets}

   \author{
        Jake K. Eager\inst{1}
        \and
        David J. Reichelt\inst{1}
        \and
        Nathan J. Mayne\inst{1}
        \and
        F. Hugo Lambert\inst{2}
        \and
        Denis E. Sergeev\inst{2}
        \and 
        Robert J. Ridgway\inst{1}
        \and
        James Manners\inst{3,4}
        \and
        Ian A. Boutle\inst{1,4}
        \and 
        Timothy M. Lenton\inst{3}
        \and 
        Krisztian Kohary\inst{1}
    }

   \institute{Physics and Astronomy, College of Engineering, Mathematics and Physical Sciences, University of Exeter, Exeter, EX4 4QL, UK
              \email{J.K.Eager@exeter.ac.uk}
         \and
             Mathematics, College of Engineering, Mathematics and Physical Sciences, University of Exeter, Exeter, EX4 4QF, UK
         \and
             Global Systems Institute, University of Exeter, Exeter, EX4 4QE, UK
         \and 
             Met Office, FitzRoy Road, Exeter, EX1 3PB, UK
    }

   \date{}
 
  \abstract{
    The majority of potentially habitable exoplanets detected orbit stars cooler than the Sun, and therefore are irradiated by a stellar spectrum peaking at longer wavelengths than that incident on Earth. Here, we present results from a set of simulations of tidally--locked terrestrial planets orbiting three different host stars to isolate the effect of the stellar spectra on the simulated climate. Specifically, we perform simulations based on TRAPPIST--1e, adopting an Earth-like atmosphere and using the UK Met Office Unified Model in an idealised `aqua--planet' configuration. Whilst holding the planetary parameters constant, including the total stellar flux (900 W/m$^2$) and orbital period (6.10 Earth days), we compare results between simulations where the stellar spectrum is that of a quiescent TRAPPIST--1, Proxima Centauri and the Sun. 
    The simulations with cooler host stars had an increased proportion of incident stellar radiation absorbed directly by the troposphere compared to the surface. This, in turn, led to an increase in the stability against convection, a reduction in overall cloud coverage on the dayside (reducing scattering), leading to warmer surface temperatures. The increased direct heating of the troposphere also led to more efficient heat transport from the dayside to the nightside and, therefore, a reduced day--night temperature contrast. We inferred that planets with an Earth--like atmosphere orbiting cooler stars had lower dayside cloud coverage, potentially allowing habitable conditions at increased orbital radii, compared to similar planets orbiting hotter stars for a given planetary rotation rate. 
    
    }

   \maketitle
%

\section{Introduction}

    Several potentially habitable terrestrial exoplanets have been detected, including Proxima Centauri b \citep{anglada16} and TRAPPIST--1e \citep{gillon17}, orbiting M--dwarf stars which are smaller and cooler than the Sun (G--dwarf). The change in the host star brightness and temperature leads to two important consequences. Firstly, for a planet to orbit in the habitable zone \citep{kasting93} around an M--dwarf, it must have a smaller orbital radius (and therefore shorter period) than that of Earth. Therefore the planet will experience stronger tidal forces from the host star, compared to the Earth from the Sun, which is likely to result in the planet's rotation rate and orbital period becoming synchronised - known as tidal locking \citep{pierrehumbert19}. Secondly, the amount of stellar radiation incident on the planet peaks at longer wavelengths, due to the lower temperature of M--dwarfs compared to G--dwarfs \citep[eg. see][]{joshi12,shields13,rushby19}. Another important difference between G and M--dwarfs is the occurrence rates and strength of stellar flares, and overall stellar activity, both being much higher in M--dwarfs \citep[see e.g.][for Proxima Centauri]{howard_2018}. This has important implications for both the atmospheric composition, for example, in terms of stratospheric ozone cycling \citep{yates20}, and the habitability of planets orbiting such stars. Initial studies have been performed in 1D \citep{tilley_2019}, but extension to 3D is required given the assumption of tidal locking for planets such as TRAPPIST--1e and Proxima Centauri b, resulting in a permanent dayside and nightside, the latter receiving no direct stellar irradiation. In this work we focus on the differences caused exclusively by the quiescent stellar spectra and reserve inclusion of stellar activity to future work.
    
    The climates of the TRAPPIST--1 planets \citep{wolf17b,turbet18,fauchez20} and Proxima Centauri b \citep{turbet16,boutle17,delgenio19,boutle20} have been simulated using different model infrastructures and exploring different facets of the climate system. The vast majority of these simulations reveal a similar dynamical structure, of a dominant, coherent zonal flow, or jet, that transports heat from the dayside to nightside. However, a direct comparison to isolate the significance of the spectrum of the host star has yet to be performed. The effect of stellar type, through differing atmospheric absorption, on cloud, convection and day--night heat and moisture transport are key in determining the impact differences in spectra between different stars will have on the planetary climate. 
    
    For terrestrial exoplanets, \citet{yang13} demonstrated that clouds produce a negative feedback extending the inner edge of the habitable zone. As the overall stellar irradiance increases, so does convection, cloud coverage and consequently the albedo on the dayside, thus cooling the planet. This is only possible if there is a large water supply on the dayside of the planet (e.g. on an aquaplanet). \citet{yang14} and \citet{koll16} employed two--box (dayside and nightside) models to determine what controls the surface temperature. \citet{yang14}, in particular, showed that the planet's nightside acts as a ``radiator fin'' allowing outgoing longwave radiation (OLR) to escape from the atmosphere, cooling the planet, due to the low level of high--altitude cloud. This is because on the dayside the water vapour and cloud greenhouse effects reduce the efficiency of the local atmosphere in radiating stellar energy to space. This energy is instead transported by the atmosphere to the nightside where there is a strong temperature inversion, and the cloud greenhouse effect is negligible/reversed, so that infrared energy is easily emitted to space. \citet{yang14} showed that when the emissivity of the nightside is increased, the dayside surface temperature decreases significantly, whereas increasing dayside emissivity leads to small increases in temperature. 
    
    \citet{boutle17} showed that for a simulation of Proxima Centauri b, vigorous convection over the sub--stellar point acted to transport heat and moisture vertically to the altitude of the zonal jet. Recently, \citet{sergeev20} have explored the differences obtained when employing various treatments and parameterisations of convection within 3D simulations of a tidally--locked terrestrial exoplanet, and performing high--resolution convection--permitting simulations free from such approximated treatments. \citet{sergeev20} showed that important differences in the vertical and horizontal transport of heat and moisture exist between coarse--resolution, employing convection parametrisations and high--resolution simulations with explicit convection. However, these studies have not yet been extended to explore the impact of differing stellar spectra on the behaviour of the convective transport, cloud coverage and day--night transport.
    
    The impact different stellar spectra have on a planetary climate has been studied for rapidly rotating planets \citep{shields13}. \citet{shields13} found that when holding the total stellar irradiance received by a planet constant, planets orbiting cooler, redder stars exhibit higher global mean surface temperatures than those orbiting warmer stars. This was due to increased direct absorption of incident stellar radiation by the atmosphere for planets orbiting cooler stars. The stellar spectra of an M--dwarf overlaps considerably more with the absorption features of CO$_2$ and H$_2$O than that of a G--dwarf, with the former emitting a larger proportion of radiation in the near infrared \citep{pierrehumbert10}. \citet{shields13} also found that the H$_2$O ice albedo feedback (where, as ice forms, more light is reflected from the planetary surface leading to further cooling and increased ice coverage) was weaker for planets orbiting cooler stars. This is due to ice albedo's wavelength dependence, which decreases with wavelength above 0.5 $\mu$m, leading to a smaller contrast between ice and water \citep{joshi12}. \citet{shields19} took this further to find that a planet orbiting an M-dwarf absorbs 12\% more incident solar energy than its G-dwarf counterpart for an Earth--like configuration with a 24 hour rotational period. Meanwhile, \cite{yang19c} found that an increase in atmospheric absorption of stellar radiation led to an increase in relative humidity at higher altitudes, globally, causing a significant decrease in OLR.
    
    In this study, we extend on previous works by investigating the effect that different stellar spectra have on the planetary climate of tidally--locked planets with Earth--like atmospheres, focusing on cooler stars around which current, potentially habitable, targets have been detected. We performed simulations using the Met Office 3D climate model, the Unified Model (UM), based on the planetary parameters for TRAPPIST--1e and a 1\,bar N$_2$ dominated atmosphere with 400 ppm CO$_2$. Further simulations were performed, replacing the stellar spectrum of TRAPPIST--1 with that of Proxima Centauri and the Sun, holding all other parameters constant, and retaining a tidally-locked configuration. Of course, setting a constant rotation rate across our experiments would not be physically consistent with tidally--locked planets obeying Kepler's laws. However, the effect of changes in the rotation rate on exoplanet climates has been well studied \citep[e.g.][]{merlis10,haqq-misra2018,penn18,komacek19} and is not our focus here. Additionally, increasing the gravity in a simulation of a given planet leads to a cooling for cases where a dilute, radiatively active condensible (such as water in our configuration) is present \citep{thomson19,yang19}. Therefore, as we look to isolate the effect that changing the stellar spectrum has on the planetary climate, we maintain a constant top--of--atmosphere incident flux, orbital period, atmospheric composition, planetary mass and radius for all our simulations.

    In \secref{sec:model} we give an overview of the UM (which has now been employed, and detailed in many exoplanet studies) and our specific configurations, followed by presenting our results in \secref{sec:results}. In \secref{subsec:temp_dyn} we explore the basic climatology of our simulations, through the surface temperature and winds. This is followed by investigation of the moisture and cloud coverage in \secref{subsec:moisture}, and separation of the radiative, advective, latent and boundary layer turbulent contributions to the heating and evaporation/condensation in \secref{subsec:balance}. Finally in \secref{sec:conclusions} we present our conclusions and discuss both the limitations of our approach, and the potential implications for the habitability of tidally--locked planets with Earth--like atmospheres. 
    We find that planets orbiting cooler stars absorb more shortwave stellar radiation directly in the troposphere, which leads to more efficient zonal circulation and a smaller temperature gradient between the day and nightside. The increase in the ratio of radiation absorbed by the atmosphere compared to the surface results in a dayside with less vigorous convection - reducing dayside cloud cover and hence the overall planetary albedo. This results in planets orbiting cooler stars being globally warmer than those orbiting hotter stars. Overall, we find that planets orbiting cooler stars have larger regions on the dayside that can support liquid water, and infer that such planets likely maintain habitable temperatures out to larger orbital radii (and lower overall incident stellar fluxes) than their counterparts orbiting hotter stars. 

\section{Model Setup}
\label{sec:model}
    
In this work we use the Met Office general circulation model (GCM), the UM, which has been adapted to a range of exoplanet applications and used for a large number of studies covering hot Jupiters \citep{mayne14a,amundsen16,helling16,mayne17,tremblin17,drummond18a,drummond18c,lines18a,lines18b,lines19,sainsbury19,debras19,debras20,drummond20}, mini--Neptunes/Super Earths \citep{drummond18b,mayne19} and terrestrial planets \citep{mayne14b,boutle17,lewis18,fauchez20,yates20,boutle20,joshi20,sergeev20}. For this work, we follow a similar configuration to that of \citet{boutle17} and \citet{lewis18}, based on the Global Atmosphere 7.0 configuration \citep{walters19}. The UM's ENDGame dynamical core uses a semi-implicit semi-Lagrangian formulation to solve the non-hydrostatic, fully compressible deep atmosphere equations of motion \citep{wood14}. Processes that occur on a scale smaller than the size of the grid boxes are parametrised. Convection uses a mass-flux approach based on \citet{gregory90}, water clouds use the PC2 scheme detailed in \citet{wilson08} incorporating mixed phase microphysics based on \citet{wilson99}, and turbulent mixing uses an approach based on \citet{lock00,brown08}. The simulations were configured as an aquaplanet, using a single layer slab homogeneous flat surface as the inner boundary (planet's surface), which is based on \citet{frierson06}. It represents an ocean surface with a 2.4 m mixed layer with a heat capacity of 10$^7$ J/K/m$^2$, with no horizontal heat transport. The emissivity of the surface is fixed at 0.985 and the albedo is spectrally dependant and varies with stellar zenith angle, based on \citet{jin11}. Sea ice formation is not considered in the model, with the surface remaining as liquid water throughout. The Suite of Community Radiative Transfer codes based on Edwards and Slingo (SOCRATES) scheme treats the radiative transfer in the UM, employing the correlated--k method. SOCRATES has been adapted and tested for a range of exoplanet configurations \citep[e.g][]{amundsen14,amundsen17}, but in this work we use a configuration similar to that used to study Earth \citep{walters19}.  Longwave ``planetary'' radiation is treated via 12 bands (between 3.3 $\mu$m-10 mm) while shortwave ``stellar'' radiation is treated by 29 bands (0.20-20 $\mu$m) with the opacity data obtained from the NASA Goddard Institute for Space Studies\footnote{From directories \url{sp_sw_dsa_ar} and \url{sp_lw_dsa_ar} at \url{https://portal.nccs.nasa.gov/GISS_modelE/ROCKE-3D/spectral_files/}}.
  
\begin{figure}
    \centering
    \includegraphics[width=\linewidth,trim=0cm 0cm 1.2cm 1cm,clip]{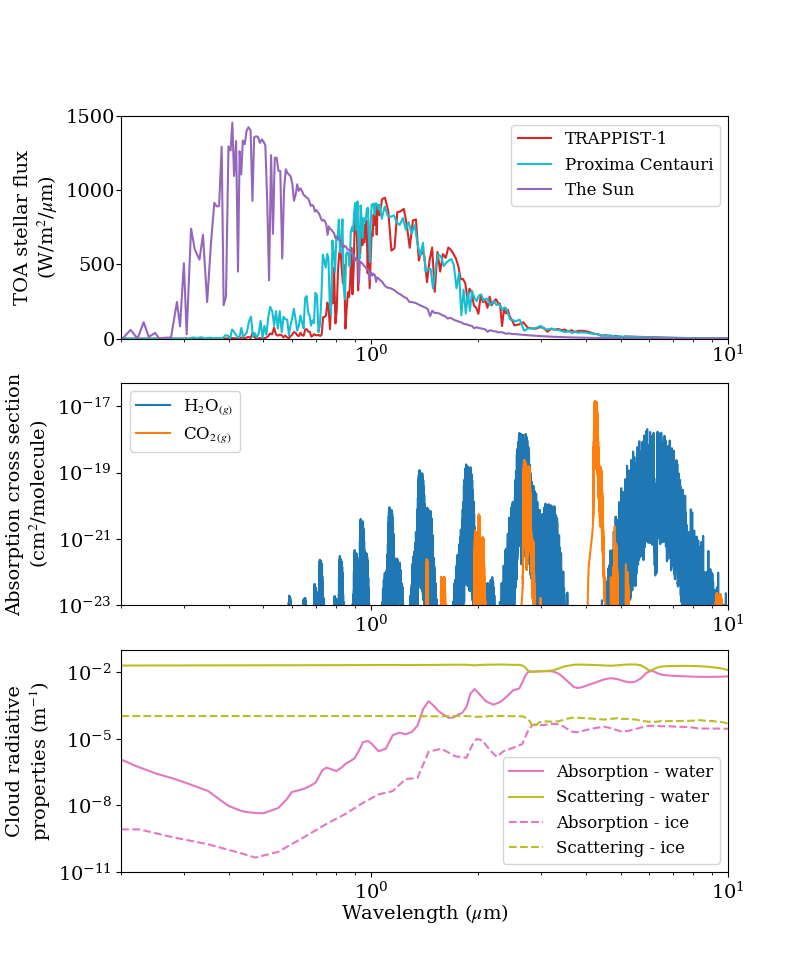}
        \caption{Wavelength vs the Stellar flux per wavelength (top), received at the top--of--atmosphere (TOA) for a planet orbiting TRAPPIST--1 (red), Proxima Centauri (cyan) and the Sun (purple), with a fixed total stellar flux of 900 W/m$^{2}$. Stellar profiles created using the BT-settl model grid of theoretical spectra \citep{rajpurohit13} with stellar parameters from \tabref{tab:stellar_params}. The middle figure shows the absorption cross section per molecule for water vapour (blue) \citep{polyansky18} and carbon dioxide gas (orange) \citep{tashkun11} against wavelength. Absorption cross sections (center) are for pressure of \textasciitilde 800\,hPa and air temperature of \textasciitilde 230\,K using the ExoMol \citep{tennyson16} database, generated using ExoCross \citep{Yurchenko18}. Also shown are the cloud absorption (pink) and scattering (green) rates for liquid water (solid) and water ice (dashed). These assume typical cloud droplet radii of 9 $\mu$m and 30 $\mu$m for liquid and ice cloud, respectively.}
         \label{fig:stellar_spectra}
    \end{figure}

As our focus is the effect that different host star emission has on the climate of a planet, we use input spectra for three different stars: TRAPPIST--1, Proxima Centauri and the Sun. The stellar parameters for these stars are shown in \tabref{tab:stellar_params}, and their spectra are shown in \figref{fig:stellar_spectra} (top), generated using the BT-settl model of theoretical spectra \citep{rajpurohit13}. \figref{fig:stellar_spectra} (middle) shows the wavelength dependence of the absorption cross section for water vapour (blue) \citep{polyansky18} and carbon dioxide (orange) \citep{tashkun11}. The absorption cross sections were generated using the ExoMol \citep{tennyson16} database, and the ExoCross software \citep{Yurchenko18}, for an atmospheric pressure and temperature of \textasciitilde 800\,hPa and \textasciitilde 230\,K, respectively. The Sun's emission peaks at visible wavelengths, whereas TRAPPIST--1 and Proxima Centauri peak in the infrared, with a larger fraction of the radiation emitted at >1 $\mu$m - the region where carbon dioxide and, particularly, water vapour begin to absorb. TRAPPIST--1 is the coolest star and emits more radiation at longer wavelengths than Proxima Centauri, for constant total flux. \figref{fig:stellar_spectra} (bottom) shows the cloud radiative properties. Scattering for both ice and liquid water cloud remains relatively constant across the stellar spectrum, and thus the cloud albedo will remain constant between the simulations for the same cloud distributions. In terms of absorption rates, both ice and liquid water clouds have global minima at the peak of the Sun's stellar spectrum at \textasciitilde 0.4 $\mu$m, while Proxima Centauri and TRAPPIST-1 peak where cloud absorption rates are about 3 orders of magnitude higher. We can thus expect that there will be increased atmospheric absorption by clouds for the cooler stars.
    
    \begin{table*}[ht]
        \caption{Stellar parameters for TRAPPIST--1 \citep{fauchez20}, Proxima Centauri \citep{schlaufman10} and the Sun as well as the semi--major axis for the planet in our simulations.}
        \label{tab:stellar_params}
        \centering
        \begin{tabular}[t]{l*{4}{>{\centering\arraybackslash}p{0.15\linewidth}}}
            \toprule
            \textbf{Host star}      &  \textbf{Effective temperature (K)} & \textbf{g (m/s$^{2}$)} & \textbf{Metallicity  (dex)} & \textbf{Semi--major axis (AU)} \\
            \midrule
            TRAPPIST--1 & 2600   & 1000 & 0 & 0.02928 \\
            Proxima Centauri & 3000 & 1000 & 0.3 & 0.04800  \\
            The Sun             & 5700  & 274 & 0.012 & 1.230 \\
            \bottomrule
        \end{tabular}
    \end{table*}
    
   As discussed, in order to isolate the impact of the different stellar spectra we perform three simulations all with the planetary parameters of TRAPPIST--1e, taken from \citet{gillon17,grimm18} and shown in \tabref{tab:planet_params}, which are consistent with those used recently by \citet{fauchez20}. The simulations use the input stellar spectra for TRAPPIST--1, Proxima Centauri and the Sun shown in \figref{fig:stellar_spectra}, termed T1:T1e, ProC:T1e and Sun:T1e, respectively. As our primary focus is investigating the effect of differing stellar spectra of our three host stars, we maintain a fixed total stellar irradiance at the planet. In practice this requires altering the orbital semi--major axis, with the values show in \tabref{tab:stellar_params}. In reality, we would of course expect the orbital period to increase with semi--major axis, due to Kepler's third law, with a commensurate change expected in the rotation rate to retain a tidally--locked configuration. However, as changes in the rotation rate lead to well studied changes in the circulation and climate \citep{merlis10,penn18}, we adopt a constant orbital period and angular frequency of rotation. The simulations are also performed at zero obliquity and eccentricity, consistent with tidal locking. It is important to note that the simulations ProC:T1e and Sun:T1e are not designed to represent any particular planet, but solely to investigate the isolated impact of the different stellar spectra.
   
    \begin{table}[ht]
        \caption{The parameters used for all planetary configurations, based on TRAPPIST--1e from \citet{gillon17,grimm18,fauchez20}.}
        \label{tab:planet_params}
        \centering
        \begin{tabular}[t]{l>{\centering\arraybackslash}p{0.3\linewidth}}
            \toprule
            \textbf{Parameter} &  \\
            \midrule
            Stellar irradiance (W/m$^{2}$) & 900   \\
            Orbital period (Earth days) & 6.10  \\
            Angular frequency (rad/s) & 1.19$\times 10^{-5}$ \\
            Eccentricity & 0 \\
            Obliquity ($\degree$) & 0 \\
            Radius (km) & 5800 \\
            Surface gravity (m/s$^{2}$) & 9.12  \\
            \bottomrule
        \end{tabular}
    \end{table}
    
    All simulations use a horizontal resolution of 2.5$\degree$ in longitude by 2$\degree$ in latitude, with 38 vertical levels between the surface ($z=0$ km) and the top--of--atmosphere ($z=40$ km). The vertical levels are quadratically stretched to enhance the resolution at the surface. All simulations ran for 8000 Earth days, with a time step of 1200 seconds, with equilibrium being reached after 1000 Earth days, as determined through stable global mean surface temperatures and balance of the top--of--atmosphere flux (not shown). The data presented in \secref{sec:results} are temporal averages from 1000 to 8000 days, and where a vertical coordinate is used the data are converted from the model height grid to $\sigma$, where $\sigma=p/p_s$ and $p$ is the pressure and $p_s$ the surface pressure for that specific model column. The global average surface pressure for all simulations is 1\,bar. The sub-stellar point, the point closest to the host star, is located at (0,0)$\degree$ and the anti-stellar point, the point furthest from the host star, is located at (0,180)$\degree$. Finally, spatial averages are also presented in \secref{sec:results}, where dayside averaged quantities include data from $-90\degree$ to 90$\degree$ in latitude and $-90\degree$ to 90$\degree$ in longitude, and nightside averaged quantities include data from $-90\degree$ to 90$\degree$ in latitude and $-180\degree$ to $-90\degree$ and 90$\degree$ to 180$\degree$ in longitude. Units given in terms of days refer to the duration of an Earth day. UM output was processed and plotted using Python's Iris \citep{iris} and Matplotlib \citep{matplotlib} packages.

\section{Results}
\label{sec:results}
   
    In this section we present results from our three simulations: T1:T1e, ProC:T1e and Sun:T1e. This begins with the basic temperature and wind structure (\secref{subsec:temp_dyn}), before moving to exploration of the moisture, cloud coverage and the subsequent effect on the radiation budget (\secref{subsec:moisture}). We finish with the components contributing to the heat and water vapour budget (\secref{subsec:balance}).
    
    \subsection{Surface Temperature and Atmospheric Dynamics}   
    \label{subsec:temp_dyn}
    
\begin{figure*}
       \centering
       \includegraphics[width=\linewidth,trim=2.5cm 0cm 2.5cm 4cm,clip]{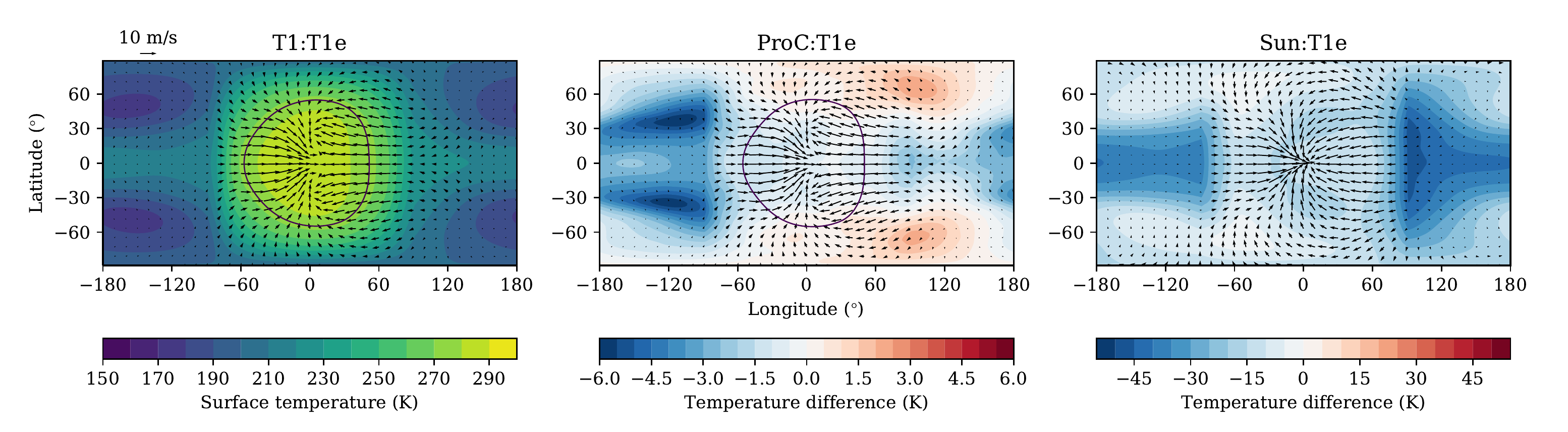}
          \caption{A map of the surface temperatures (colour scale) for the T1:T1e simulation (left). The following two plots show the difference in surface temperature from T1:T1e for the ProC:T1e (middle) and the Sun:T1e (right) cases. A negative difference (blue) indicates a cooler surface than T1:T1e. Near surface (10m) wind vectors (arrows) are also shown on each plot. The sub-stellar point is located at (0$\degree$,0$\degree$). A contour (black) is shown for the 273\,K surface isotherm, but this temperature is not reached over an extended region for the Sun:T1e case. Note: the difference in colour scale between ProC:T1e and Sun:T1e. Only the temperature field is subtracted, the winds are the unaltered values for each simulation.
                  }
             \label{fig:T_surf}
       \end{figure*}

    A natural metric to describe the basic climatic state is the surface temperature. \figref{fig:T_surf} shows the surface temperature variation across latitude and longitude for our three simulations, with the winds at 10m shown as vector arrows. The left figure shows the absolute surface temperature for the T1:T1e case, differences are then shown via a subtraction of the T1:T1e temperature field from either the ProC:T1e or Sun:T1e results as the middle, and right panels, respectively. \figref{fig:T_surf} shows that as the temperature of the host star increases (left to right) the planetary surface temperature generally decreases. The greatest cooling is seen on the nightside, predominantly at the equator, with some warming in the polar regions of the ProC:T1e case. This suggests there may be an asymmetry between the changes in the meridional and zonal transport efficiency. The region of the surface above 273\,K is enclosed by the black contour in \figref{fig:T_surf}. These are similar in magnitude between the T1:T1e and ProC:T1e cases, however the Sun:T1e case does not have a substantial region of the planetary surface that may sustain liquid water and may be considered less habitable as a result. All simulations have similar near surface winds, showing a convergence towards the sub-stellar point, due to solar forcing giving rise to a region of intense convection as discussed in \citet{boutle17,sergeev20}.
    
    \tabref{tab:T_surf} shows the spatial average dayside, nightside and global surface temperatures for the three simulations. The values in \tabref{tab:T_surf} confirm that the simulation with the coolest star, T1:T1e, is the warmest, with the Sun:T1e case exhibiting the coldest temperatures. The day-night temperature contrast is smallest for the T1:T1e simulation, suggesting the most efficient day-night circulation of the three simulations, while the Sun:T1e case has the largest contrast and the weakest circulation. The T1:T1e and ProC:T1e cases have similar temperatures, with T1:T1e consistently warmer on the order of 1\,K. The small differences in stellar spectra between TRAPPIST-1 and Proxima Centauri (\figref{fig:stellar_spectra}) may have a small effect on planetary climate, which is only amplified by larger contrasts in effective stellar temperature.
        
\begin{figure*}
       \centering
       \includegraphics[width=\linewidth,trim=2.5cm 0cm 2.5cm 0cm,clip]{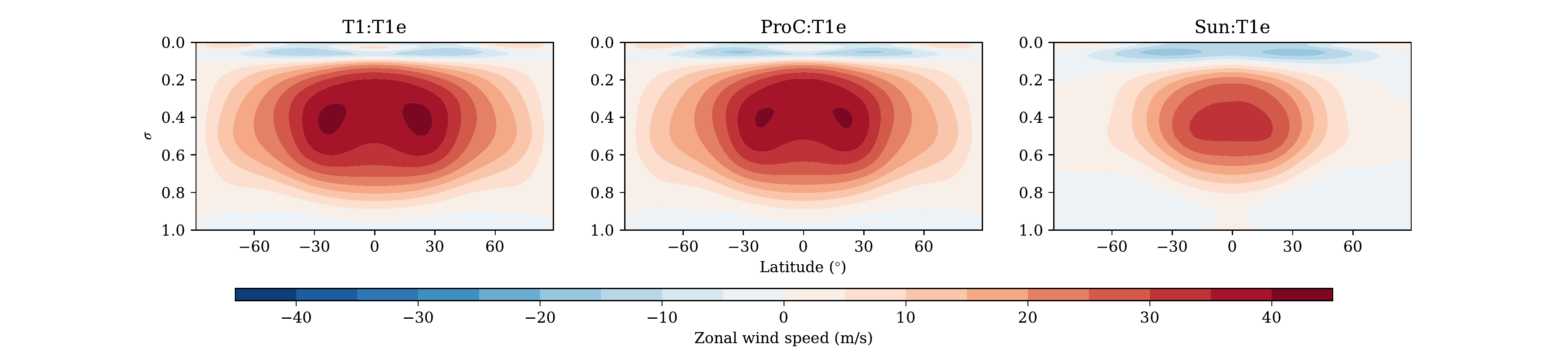}
          \caption{Latitude vs $\sigma$ (pressure divided by surface pressure), showing the zonal wind speed (colour scale), taken as a longitudinal average, for the T1:T1e (left), ProC:T1e (middle) and Sun:T1e (right) simulations. Positive values for zonal wind (red) represent eastward flow.
                  }
             \label{fig:zonal_wind}
       \end{figure*}

    The dominant component of the heat redistribution from the day to night side of the planet, is the zonal jet \citep[e.g.][]{lewis18}. \figref{fig:zonal_wind} shows the longitudinal (and temporally) averaged zonal wind for latitude against $\sigma$, shown for the T1:T1e (left), ProC:T1e (middle) and Sun:T1e (right) simulations. The super--rotating equatorial jet (in red) reduces in magnitude as the host star increases in temperature (left to right). As shown by \citet{showman10,showman11}, the zonal jet is accelerated via large--scale wave patterns which are driven by the day--night temperature contrast, and further shaped by the vertical and latitudinal heating gradients. \citet{lewis18} also showed that changes in the radiative properties of the surface, i.e. moving from bare land to ocean, resulted in a change in the temperature structure and, thereby, the jet acceleration. In our simulations, as we move from hotter to cooler host stars there is an increase in the overall absorption of radiation on the dayside (see \secref{subsec:moisture}). One might expect this to result in a larger day--night temperature contrast for cooler stars, as opposed to the reduction shown in \tabref{tab:T_surf}. However, as the absorption is dominated by the atmosphere (as opposed to the surface), this results in a day--night contrast extending over a larger range of pressures, i.e. higher into the atmosphere for cooler stars (see \secref{subsec:moisture}). We speculate that this acts to extend the vertical region over which momentum convergence acts to accelerate the jet and, indeed, the jet structure persists over a broader vertical (and meridional) range for the cooler star simulations, as shown in \figref{fig:zonal_wind}. The vertical component of momentum convergence has been shown to be vital for accelerating super--rotating equatorial flows \citep{showman11} in hot Jupiters, and we have studied the detailed wave responses in these cases \citep{debras19,debras20}. However, we reserve such a detailed study of these simulations for future work, and here simply note that the jet is stronger for planets orbiting cooler stars, and the flow acts to transport heat and, critically, moisture zonally around the planet. Planets orbiting cooler stars can also have a stronger nightside equatorial return flow near the surface, as seen in \figref{fig:T_surf}. 
    
    \begin{table}[ht]
        \caption{Mean surface temperatures for global, dayside, nightside and the temperature contrast (nightside subtracted from dayside), for T1:T1e, ProC:T1e and Sun:T1e.}
        \label{tab:T_surf}
        \centering
        \begin{tabular}[t]{l*{4}{>{\centering\arraybackslash}p{0.125\linewidth}}}
           \toprule
          \textbf{Simulation} &  \multicolumn{4}{c}{\textbf{Temperature (K)}} \\
           & \textbf{Global} & \textbf{Dayside} & \textbf{Nightside} & \textbf{Contrast} \\
           \midrule
           T1:T1e & 231.2 & 260.8 & 201.6 & 59.2 \\
           ProC:T1e & 229.8 & 260.1 & 199.5 & 60.6  \\
           Sun:T1e & 209.4 & 245.5  & 173.4 & 72.1 \\
           \bottomrule
       \end{tabular}
    \end{table}
   
 \subsection{Moisture and Cloud in the Atmosphere}   
    \label{subsec:moisture}
    
\begin{figure*}
       \centering
       \includegraphics[width=\linewidth,trim=2.5cm 1cm 2.5cm 3.5cm,clip]{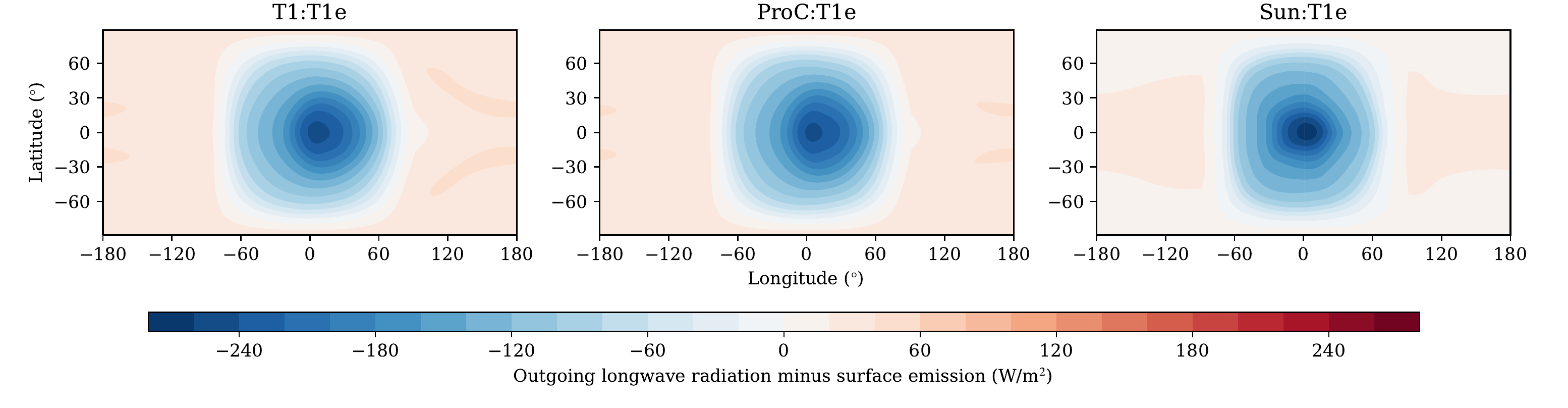}
          \caption{A map of the outgoing longwave radiation (OLR) minus the planetary surface emission as a colour scale for T1:T1e (left), ProC:T1e (middle) and Sun:T1e (right) simulations. A positive difference (red) indicates an increase in OLR emission relative to the surface. The sub-stellar point is located at (0$\degree$,0$\degree$).
                  }
             \label{fig:OLR}
       \end{figure*}

    Water vapour and cloud play an important role in the radiation budget, particularly in shaping the outgoing longwave radiation (OLR) and determining the contributions of the atmosphere compared to the planetary surface. \figref{fig:OLR} shows the OLR for our three simulations after subtraction of the longwave surface emission. 
  
    All the simulations show the same pattern of dayside OLR originating from colder levels in the atmosphere than the surface, due to high--altitude clouds and water vapour. The nightside OLR indicates emission from warmer levels than the surface due to cloud and water vapour around the nightside temperature inversion and the lack of high-altitude cloud.
    In the rest of this section we investigate the changes in moisture and cloud coverage and use this to understand the changes in radiation emission between the simulations.
    
       \begin{figure*}
       \centering
       \includegraphics[width=\linewidth,trim=2.5cm 0cm 2.5cm 4cm,clip]{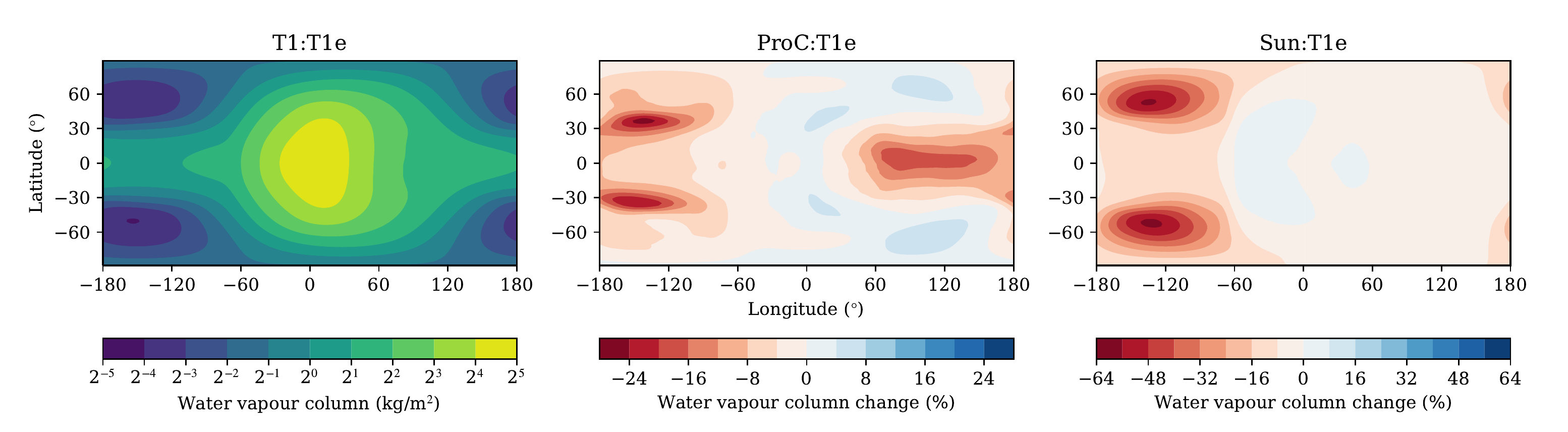}
          \caption{A map of the column integrated water vapour (mass of water per cross section area, colour scale) for the T1:T1e simulation (left). The following two plots show the change in water vapour column content for ProC:T1e (middle) and Sun:T1e (right) with the equivalent water vapour column content of T1:T1e, if it were at the same temperature profile of ProC:T1e and Sun:T1e, respectively at the same relative humidity as T1:T1e. The percentage is calculated from the total water vapour content of the column. A positive difference (blue) indicates a moister column than the T1:T1e, removing the effect of the Clausius--Clapeyron relation of q with temperature. The sub-stellar point is located at (0$\degree$,0$\degree$). Note the difference in scale between ProC:T1e and Sun:T1e.
                  }
             \label{fig:q_col}
       \end{figure*}
    
       \begin{figure*}
       \centering
       \includegraphics[width=0.8\linewidth,trim=0cm 0cm 0cm 0cm,clip]{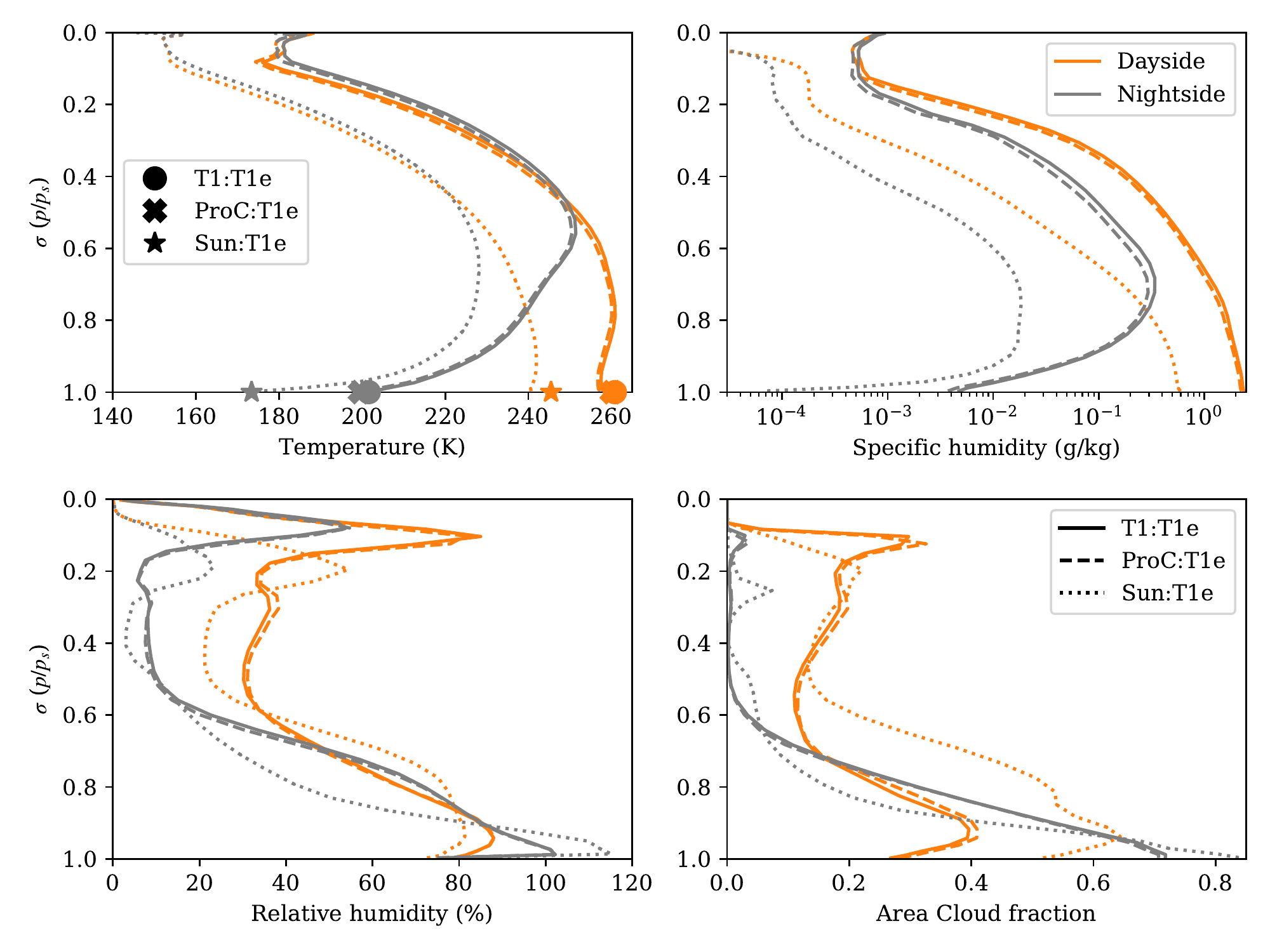}
          \caption{Temperature (top left), specific humidity (top right), relative humidity (bottom left) and area cloud fraction (bottom right) horizontally averaged over the dayside (orange) and nightside (grey) hemispheres, plotted against $\sigma$ (pressure divided by surface pressure). These are shown for all three simulations: T1:T1e (solid line), ProC:T1e (dashed line) and Sun:T1e (dotted line). The left figure also includes the hemisphere averaged surface temperatures from \tabref{tab:T_surf} as a \ding{108} for T1:T1e, a \ding{54} for ProC:T1e and a \ding{72} for Sun:T1e.
                  }
             \label{fig:theta_RH_cloud}
       \end{figure*}

     Moisture transport from the dayside to the nightside of a tidally--locked planet is important, due to its effect on the OLR, both directly or through subsequent cloud formation \citep{yang14}. Generally, moisture is transported upward from the surface, via convection, in the sub--stellar region also forming cloud, with the zonal jet transporting moisture (and cloud) horizontally high--up in the troposphere, with subsidence and further condensation occurring on the nightside \citep{yang14,boutle17,lewis18,sergeev20}. To explore this for our simulations, \figref{fig:q_col} shows the column integrated water content as a function of latitude and longitude. The left panel shows the absolute water vapour column content for the T1:T1e simulation, while the middle and right panels show the percentage change in water vapour column content for the ProC:T1e and Sun:T1e simulations after subtraction of an equivalent T1:T1e water vapour content. However, as the dominant factor in the moisture content variation is due to the temperature change through the Clausius--Clapeyron relation, we have attempted to remove this component. This was done by calculating the difference in relative humidity of the ProC:T1e and Sun:T1e cases from the T1:T1e case, and integrating the equivalent water column content as a percentage change from the ProC:T1e and Sun:T1e cases, respectively. In \figref{fig:q_col} we see that as the host star temperature increases (left to right) the nightside of the planets get relatively drier, beyond the drying through the Clausius--Clapeyron relation i.e. a general decrease in the relative humidity of the atmospheric column, suggesting a decrease in the atmospheric transport observed in the zonal jets in \figref{fig:zonal_wind}.
     
    The advection of heat, moisture and cloud from the dayside results in a nightside temperature inversion \citep{joshi20} leading to the radiator fin effect discussed in \citet{yang14}, whose magnitude depends on the opacity on the nightside determined by the water vapour and cloud content. From \figref{fig:q_col}, we expect the planets orbiting cooler stars to have a stronger water vapour greenhouse effect on the dayside, but increase the effect of the nightside radiator fin by increasing cloud content and, hence, the OLR. This can be explored further by using vertical profiles of the temperature, moisture content and cloud fraction from our simulations. \figref{fig:theta_RH_cloud} shows the hemispherically averaged variation with $\sigma$ of air temperature (top left), specific humidity (top right), relative humidity (bottom left) and area cloud fraction (bottom right). Area cloud fraction is the area within a model grid box which is covered by cloud.    
    
    Firstly, \figref{fig:theta_RH_cloud} shows a clear temperature inversion on the nightside of all simulations (top left), linked to the circulation in the free atmosphere and radiative cooling of the surface. Additionally, the day--night temperature difference between the vertical profiles (top left) is also smaller for the cases with cooler stars below the inversion, consistent with the efficiency of the day--night redistribution also increasing toward cooler host stars.
    
    On the dayside, with the majority of the atmosphere cooler than the surface (top left of \figref{fig:theta_RH_cloud}), greenhouse gases and clouds decrease the OLR relative to surface emission (shown in \figref{fig:OLR}). The hemispherically averaged specific humidity is highest at all levels on the dayside and nightside for the T1:T1e case and lowest for the Sun:T1e case (top right of \figref{fig:theta_RH_cloud}). We would therefore expect a stronger greenhouse effect for the T1:T1e simulation. For all simulations, on the nightside the combination of the temperature inversion, warmer air temperatures and stronger zonal transport leads to an increased water vapour and a higher OLR relative to the surface. T1:T1e is the moistest of our simulations resulting in the largest increase in OLR due to greenhouse gases, closely followed by ProC:T1e.
    
    The relative humidity (bottom left of \figref{fig:theta_RH_cloud}) shows an increase at high altitudes, which is both larger in magnitude and higher in the atmosphere for the T1:T1e and ProC:T1e simulations compared to the Sun:T1e case. This effect has already been noted by \cite{yang19c}, who demonstrated that the increased high--altitude relative humidity for planets absorbing more shortwave radiation in the atmosphere resulted in an increased water vapour greenhouse effect and, therefore, a reduction in OLR. \tabref{tab:radiative_component_forcing} shows the TOA radiative effects of water vapour and clouds for all our simulations. The water vapour and cloud radiative effects were isolated via extra `diagnostic' radiative transfer calculations, which did not affect the model evolution. One calculation omitted only the water vapour opacity, and a second calculation (termed `clear--sky') omitted the cloud radiative effects, both of which can then be compared to the baseline simulation for all cases to provide the values in \tabref{tab:radiative_component_forcing}. For our simulations an increased water vapour greenhouse effect for planets orbiting cooler stars is also found \citep[as in][]{yang19c}, shown in \tabref{tab:radiative_component_forcing}, top rows, column three, but a clear and commensurate change in the OLR is not present in \figref{fig:OLR}, due to the greater contribution of the cloud coverage to the dayside greenhouse effect shown in the same table, bottom rows, column three.
    
    \begin{table*}[ht]
        \caption{
        The hemisphere averaged top--of--atmosphere (TOA) radiative effect in the shortwave (dayside only) and longwave (dayside and nightside) for both water vapour (top rows) and cloud (bottom rows), including the net value (sum of three separate terms divided by two) for the T1:T1e, ProC:T1e and Sun:T1e simulations. A positive radiative effect indicates a decrease in outgoing radiation. Parentheses on the longwave TOA radiative effect are the absolute percentages of the averaged TOA outgoing longwave flux for that hemisphere. The effects of the two components are isolated using `diagnostic' calculations of the radiative transfer omitting their opacities (see text for explanation).}
        \label{tab:radiative_component_forcing}
        \centering
        \begin{tabular}[t]{l*{5}{>{\centering\arraybackslash}p{0.14\linewidth}}}
            \toprule
            \textbf{Simulation} & \multicolumn{4}{c}{\textbf{TOA radiative effect (W/m$^2$)}}\\
            &\textbf{Shortwave}&\multicolumn{2}{c}{\textbf{Longwave}}&\textbf{Net}\\
            &\textbf{Dayside}&\textbf{Dayside}&\textbf{Nightside}\\
            \midrule
            \multicolumn{5}{c}{Water Vapour}\\
            \midrule
            T1:T1e & +28.4 & +9.49 (5.06\%) & -20.9 (16.1\%) & +8.49 \\
            ProC:T1e & +29.1 & +8.58 (4.67\%) & -21.5 (17.2\%) & +8.07 \\
            Sun:T1e & +5.35 & +0.704 (0.533\%) & -15.1 (20.7\%) & -4.50 \\
            \midrule
            \multicolumn{5}{c}{Cloud}\\
            \midrule
            T1:T1e & -110 & +27.0 (14.4\%)  & -4.58 (3.53\%) & -43.8 \\
            ProC:T1e & -117 & +28.2 (15.3\%) & -4.07 (3.25\%) & -46.4 \\
            Sun:T1e & -189 & +40.7 (30.8\%)& -0.822 (1.13\%)  & -74.5 \\
            \bottomrule
        \end{tabular}
    \end{table*}
    
    \figref{fig:theta_RH_cloud} also shows that the dayside averaged cloud coverage (bottom right) is largest for the warmer star, with Sun:T1e having a \textasciitilde 60$\%$ larger peak than T1:T1e and ProC:T1e, at around $\sigma=0.9$. Cloud coverage on the dayside can cool the planet by increasing the top--of--atmosphere (TOA) albedo. This is demonstrated in \tabref{tab:albedo_rad_budget} which shows the dayside albedo and total shortwave absorption as proportions of the total TOA incident stellar flux, in particular the second column.
    
    \begin{table*}[ht]
        \caption{Dayside shortwave radiation budget hemispherically averaged for the top--of--atmosphere (TOA) albedo and the dayside shortwave radiation absorption (as a fraction of the total TOA incoming shortwave radiation) for the T1:T1e, ProC:T1e and Sun:T1e simulations. The dayside albedo has been decomposed into an atmospheric and a surface contribution following \citet{donohoe11}. The dayside shortwave radiation absorption is shown for the atmosphere and surface, with the former decomposed into cloud and water vapour contribution by comparing the baseline model to the calculations where these radiative effects have been omitted (see text for explanation).
        }
        \label{tab:albedo_rad_budget}
        \centering
        \begin{tabular}[t]{l*{8}{>{\centering\arraybackslash}p{0.09\linewidth}}}
            \toprule
            \textbf{Simulation} & \multicolumn{3}{c}{\textbf{Dayside TOA albedo ($\%$)}} & \multicolumn{4}{c}{\textbf{Dayside shortwave absorption ($\%$)}} \\
            & \textbf{Total} & \textbf{Atmosphere} & \textbf{Surface} & \textbf{Atmosphere} & \textbf{Cloud} & \textbf{Water Vapour} & \textbf{Surface} \\
            \midrule
            T1:T1e & 28.8 & 28.0 & 0.767 & 43.0 & 12.2 & 15.5 & 28.3 \\
            ProC:T1e & 30.6 & 29.8 & 0.790 & 41.1 & 11.0 & 15.3 & 28.3 \\
            Sun:T1e & 54.8 & 53.6 & 1.24 & 11.5 & 4.38 & 2.77 & 33.7 \\
            \bottomrule
        \end{tabular}
    \end{table*}
    
    The total albedo increases for hotter stars, see \tabref{tab:albedo_rad_budget} column two to four, due to an increase in the albedo of the surface but is predominantly caused by increased cloud coverage, as seen in \tabref{tab:radiative_component_forcing}, bottom rows, column two. \tabref{tab:albedo_rad_budget} shows that the albedo is largest for the Sun:T1e simulation and smallest for the T1:T1e case, which has the lowest dayside cloud coverage (clouds are equally reflective in each case, see \figref{fig:stellar_spectra}). This is the dominant cause for the decrease in surface temperatures in \figref{fig:T_surf} and air temperatures, with the Sun:T1e case \textasciitilde15-20\,K cooler for all $\sigma$ on the dayside in \figref{fig:theta_RH_cloud} (top left grey). Cloud also affects the OLR budget, which can be seen in the cloud radiative effect in \tabref{tab:radiative_component_forcing}, bottom rows, columns three and four. On the dayside, cloud increases the longwave radiation retained by the atmosphere, hence decreasing OLR. However, this effect is about four times smaller than the shortwave cloud radiative effect, which is the dominant factor in the overall decrease in planetary temperature for hotter stars.
    
    \begin{table}[ht]
        \caption{The outgoing longwave radiation (OLR) budget for the dayside and nightside as a percentage of the non--reflected shortwave radiation absorbed by the planet, shown for the T1:T1e, ProC:T1e and Sun:T1e simulations.
        }
        \label{tab:OLR_budget}
        \centering
        \begin{tabular}[t]{l*{3}{>{\centering\arraybackslash}p{0.2\linewidth}}}
            \toprule
            \textbf{Simulation} & \multicolumn{2}{c}{\textbf{OLR budget ($\%$)}}  \\
             & \textbf{Dayside} & \textbf{Nightside} \\
            \midrule
            T1:T1e & 59.1 & 40.9 \\
            ProC:T1e & 59.5 & 40.5 \\
            Sun:T1e & 64.5 & 35.5 \\
            \bottomrule
        \end{tabular}
    \end{table}
    On the nightside, OLR is increased through cloud radiative effects. Near the surface, there is more cloud for the Sun:T1e simulation than the two other cases (\figref{fig:theta_RH_cloud}, bottom right). Near surface cloud has a smaller effect on cloud radiative effect as the cloud temperature is more similar to that of the surface compared to the rest of the temperatures below the inversion maxima. Due to the nightside temperature inversion, the atmosphere at $\sigma$ > 0.2 for all simulations is warmer than the planetary surface and thus radiating heat to space more efficiently, increasing the cloud radiative effect and cooling the planet. The T1:T1e and ProC:T1e cases both have more cloud between 0.6 < $\sigma$ < 0.9, compared to the Sun:T1e case, which leads to an increase in the nightside OLR relative to the clear--sky case, shown in \tabref{tab:radiative_component_forcing} (bottom rows, column four). The radiator fin effect is stronger, for planets with more efficient day-night circulation, due to an increased nightside cloud and water vapour content. \tabref{tab:OLR_budget} shows the TOA outgoing radiation budget as the dayside and nightside OLR as a proportion of the non–reflected shortwave radiation. \tabref{tab:OLR_budget} demonstrates an increase in the proportion of total radiation emitted by the planet coming from the nightside for cooler stars. This might suggest that planets orbiting cooler stars, which we have shown have generally more efficient circulation, would overall be cooler. However, our simulations show the reverse, where the cooler host star results in an overall warmer planetary climate, showing that the changes in dayside cloud albedo are the dominant mechanism \citep{yang13}. This can be seen clearly in \tabref{tab:radiative_component_forcing}, bottom rows, where the shortwave dayside cloud radiative effect is largest and dominates the net cloud radiative effect, which also increases with host star temperature, leading to the largest planetary cooling.
    
    \tabref{tab:albedo_rad_budget} shows that the shortwave reflection (albedo) on the dayside is largest for Sun:T1e and smallest for the T1:T1e simulation. \cite{donohoe11} was used to determine the atmospheric and surface contributions to the TOA dayside albedo, with the atmosphere as the dominant contribution at \textasciitilde$97.5\%$ for all simulations. The surface contribution has been significantly attenuated by the atmosphere, reducing the surface albedo by \textasciitilde$90\%$ of the actual value for all the simulations. The majority of this atmospheric albedo is produced via cloud scattering, which is the dominant contribution to TOA shortwave cloud radiative effect compared to cloud absorption (\tabref{tab:radiative_component_forcing}, bottom rows, column two). The remaining outgoing radiation budget, emitted as longwave radiation, may be distributed between dayside and nightside emission and is shown in \tabref{tab:OLR_budget}. The proportion of radiation remaining increases in favour of emission on the dayside for planets orbiting hotter stars, as demonstrated by the increased day--night surface temperature contrast, seen in \tabref{tab:T_surf}. This occurs even with an increase in cloud suppressing longwave emission on the dayside (\tabref{tab:radiative_component_forcing}, bottom rows, column three). The water vapour greenhouse effect has the opposite effect, decreasing with hotter stars (\tabref{tab:radiative_component_forcing}, top rows, column three), but has a smaller effect compared to cloud.
    
    On the nightside, both cloud and water vapour increase the nightside OLR emission due to the temperature inversion (\tabref{tab:radiative_component_forcing}, column four) enhancing the radiator fin effect. For water vapour this effect in terms of the total radiation budget decreases for cooler stars, but when compared to the total nightside OLR (parentheses) it increases with host star temperature. For cloud, this decreases for both interpretations and their combined effects contribute \textasciitilde$20\%$ of the nightside OLR. The nightside radiator fin effect is thus dominated by the day--night temperature contrast of the surface, rather than the overall cloud/water vapour structure in the atmosphere, which maintains a similar contribution to the total nightside OLR.
    
    The global net TOA water vapour radiative effect (\tabref{tab:radiative_component_forcing}, column five) is an order of magnitude smaller than the net cloud radiative effect and changes sign between the M--dwarf and G--dwarf orbiting simulations. \tabref{tab:radiative_component_forcing} shows that in T1:T1e and ProC:T1e water vapour has a net warming effect on the global budget, while in the Sun:T1e case, water vapour has a net cooling effect. The difference is mainly attributed to the decrease in shortwave absorption (\tabref{tab:radiative_component_forcing}, column two) for hotter stars, but also the decrease in water vapour greenhouse effect stemming from the decrease in moisture in the upper atmosphere on the dayside (\figref{fig:theta_RH_cloud}).

    \subsection{Heat and Moisture Budgets}   
    \label{subsec:balance}
    
       \begin{figure*}
       \centering
       \includegraphics[width=\linewidth,trim=2cm 0cm 2cm 0cm,clip]{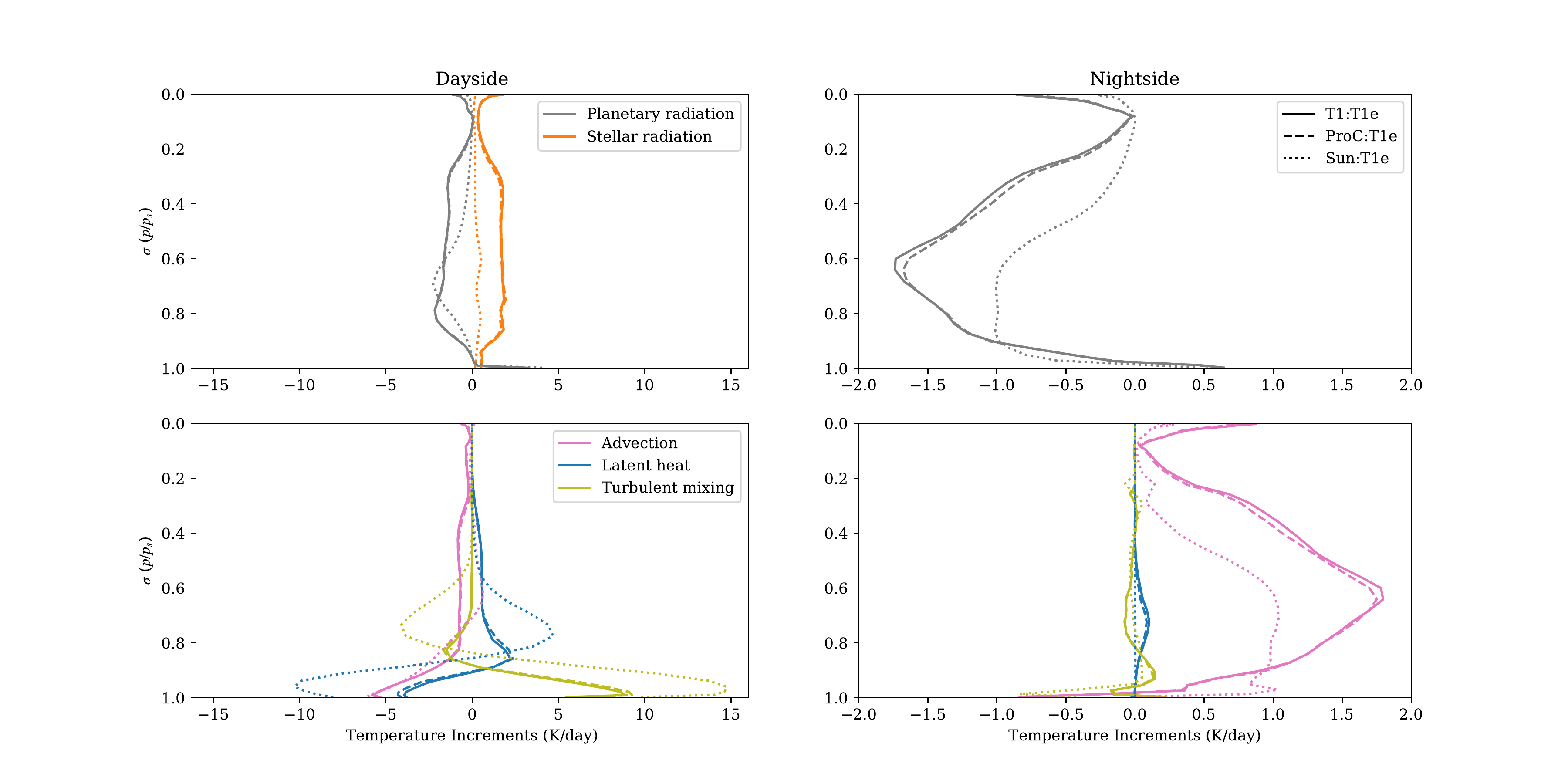}
          \caption{The rate of change of temperature, or heating profiles, known as temperature increments, plotted against $\sigma$ (pressure divided by surface pressure), for the T1:T1e (solid lines), ProC:T1e (dashed lines) and Sun:T1e (dotted lines) simulations, for each component process. The processes shown are atmospheric absorption of stellar radiation (orange, top figures), thermal emission/absorption of planetary radiation (grey, top figures), large scale circulation/advection (pink, bottom figures), latent heating/cooling of water (blue, bottom figures) and turbulent mixing (green, bottom figures). The day and nightside hemispherically averaged values are shown in the left and right figures, respectively. Note: the different x--axis limits and in equilibrium, the net heating is zero. 
                  }
             \label{fig:temp_incrs}
       \end{figure*}
    
       \begin{figure*}
       \centering
       \includegraphics[width=\linewidth,trim=2cm 0cm 2cm 0cm,clip]{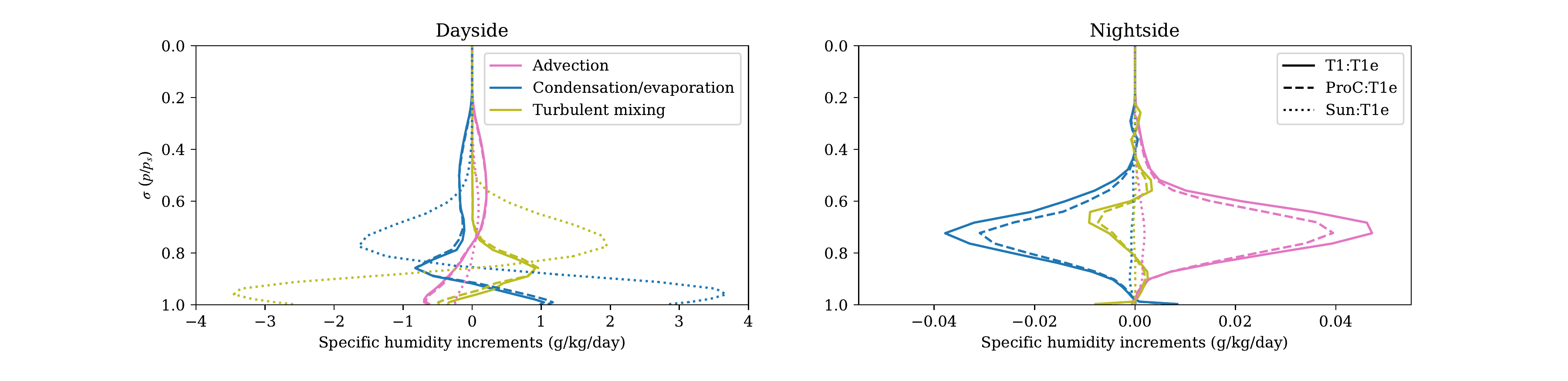}
          \caption{The rate of change of specific humidity, known as specific humidity increments, plotted against $\sigma$ (pressure divided by surface pressure), for the T1:T1e (solid lines), ProC:T1e (dashed lines) and Sun:T1e (dotted lines) simulations, for each component process. The processes shown are the large scale circulation/advection (pink), condensation/evaporation (blue) and turbulent mixing (green).  The day and nightside hemispherically averaged values are shown in the left and right figures, respectively. Note: the different x--axis limits and in equilibrium, the net specific humidity change is zero.
                  }
             \label{fig:q_incrs}
       \end{figure*}
    
    To further isolate the key or dominant processes we have separated the various contributions to the atmospheric temperature and specific humidity changes. \figref{fig:temp_incrs} shows the hemisphere averaged heating rates, or temperature increments as a function of $\sigma$, for the dayside (left figures) and nightside (right figures). For the dayside radiation (top left, \figref{fig:temp_incrs}), direct stellar radiation heats the atmosphere of the T1:T1e and ProC:T1e cases, predominantly, for 0.2 < $\sigma$ < 0.9, at \textasciitilde 2\,K/day, with the stellar heating of the atmosphere significantly reduced in the Sun:T1e case. For the T1:T1e and ProC:T1e simulations the region where $\sigma$ < 0.7 is close to radiative equilibrium (planetary radiation balances with stellar radiation), which is not the case for Sun:T1e until much higher in the atmosphere, $\sigma$ $\lesssim$ 0.3. The atmospheric absorption of stellar radiation is further quantified in \tabref{tab:albedo_rad_budget}, which shows the dayside atmospheric absorption as a proportion of the incident TOA shortwave radiation. This is nearly 4 times larger for the T1:T1e compared to the Sun:T1e case, with the T1:T1e atmosphere absorbing 1.9 $\%$ more stellar radiation than the ProC:T1e case.
    
    The direct heating of the mid to upper troposphere by cooler stars leads to an increase in convective stability in the T1:T1e and ProC:T1e cases, reducing the vertical transport of moisture, and thus, the height and magnitude of the latent heating term on the dayside (bottom left, \figref{fig:temp_incrs}). This is supported by the dayside cloud coverage shown in \figref{fig:theta_RH_cloud}, with the Sun:T1e case exhibiting more cloud at $\sigma$ > 0.5 than both the T1:T1e and ProC:T1e cases. As a result of a reduced atmospheric absorption of non--reflected shortwave radiation, the Sun:T1e case experiences an increased proportion of (non--reflected) stellar radiation absorbed at the surface on the dayside (\tabref{tab:albedo_rad_budget}, compare columns five and eight), compared to the simulations with cooler host stars. This leads to a larger turbulent flux heating the boundary layer, which is balanced by latent cooling from evaporation of precipitation falling from increased cloud and advective cooling (bottom left, \figref{fig:temp_incrs}). The shortwave atmospheric absorption has been isolated for both cloud and water vapour, in \tabref{tab:albedo_rad_budget} (column six and seven, respectively). Although the effect of both are of a similar order of magnitude for each simulation, for planets orbiting M-dwarfs, water vapour has a larger contribution to atmospheric absorption than clouds, while the opposite is true for G--dwarfs. The remaining contribution to shortwave atmospheric absorption is carbon dioxide, which is held at a constant concentration in our simulations.
    
    On the nightside (right figures, \figref{fig:temp_incrs}), advective heating is balanced by cooling via the planetary radiation emission. Advective heating comes from transport of heat from the dayside, producing the temperature inversions seen in \figref{fig:theta_RH_cloud}, and is largest for the T1:T1e case, which follows from the stronger equatorial jets seen for planets orbiting cooler stars (\figref{fig:zonal_wind}). The relatively dry atmosphere on the nightside, and lack of stellar heating at the surface, result in small latent heating and boundary layer contributions (bottom right, \figref{fig:temp_incrs}).
     
    Similarly to the temperature increments, we have isolated the contributions from different physical processes to the moisture budget in our simulations. \figref{fig:q_incrs} shows the rate of change of specific humidity in a similar format to \figref{fig:temp_incrs}. On the dayside (left, \figref{fig:q_incrs}), turbulent mixing transports water vapour from near the surface to the mid--troposphere (0.5 < $\sigma$ < 0.85), where it condenses, precipitates and then evaporates again in the boundary layer ($\sigma$ > 0.85). The large scale transport is strongest for the T1:T1e case, which can be seen in the specific humidity increment due to advection. Cool dry air is advected returning from the nightside near the surface and moist warm air is transported vertically, reducing moisture near the surface. The T1:T1e simulation has the largest nightside advection of water vapour (right, \figref{fig:q_incrs}), with the smallest found in the Sun:T1e case. On the dayside, advection reduces specific humidity near the surface. The minima for the specific humidity increment due to advection is largest for the T1:T1e and ProC:T1e cases, potentially due to a stronger return flow from the nightside. Latent and boundary layer effects occur deeper in the atmosphere for the Sun:T1e case, compared to the remaining simulations, suggesting convection becomes deeper for hotter host stars. On the nightside (right, \figref{fig:q_incrs}), moist air is transported in, by advection from the dayside, where it condenses to form nightside cloud. The cloud forms mainly around $\sigma$ = 0.7, where it descends to near the surface, as seen in \figref{fig:theta_RH_cloud} where cloud area fraction is highest there.
   
    Our results have isolated the effect that different host star spectra have on the simulated planetary climate of a tidally--locked, terrestrial exoplanet, with a modern day Earth-like atmosphere. With all else held constant, a planet orbiting a cooler star like TRAPPIST--1 or Proxima Centauri absorbs more radiation directly in the atmosphere, compared with a planet orbiting the Sun, similar to the results of \citet{shields13} for rapidly rotating non--tidally--locked planets. Increased atmospheric stellar radiation absorption leads to a decrease in the proportion of radiation absorbed by the planetary surface and an increase in static stability and decrease in convection, leading to reduced dayside cloud coverage. This decreases the albedo, and leads to a warmer planet, with the T1:T1e simulation globally 1.4\,K and 21.8\,K warmer than the ProC:T1e and Sun:T1e cases, respectively. The day--night temperature and atmospheric moisture content contrast is also smallest for the T1:T1e case. 
    
\section{Conclusions}
\label{sec:conclusions}

    We have used the Met Office 3D GCM to compare simulations of the climates of a planet orbiting three different host stars, two of which are M-dwarfs known to have near Earth sized planets orbiting in their habitable zone, with the third being the Sun, a G-dwarf. We assume an Earth--like atmospheric composition, and a tidally-locked state. With the stellar irradiance, and other planetary parameters held constant, planets orbiting cooler stars experience an increased proportion of incident radiation absorbed directly by the troposphere, compared to the surface. This is due to the increase in cloud, water vapour and carbon dioxide's ability to absorb stellar radiation when orbiting cooler stars. This leads to an atmosphere that is more statically stable, reducing dayside convection and thus cloud coverage, compared to hotter stars. For these planets orbiting hotter stars, increasing cloud coverage increases the planetary albedo, decreasing the overall proportion of radiation absorbed by the planet, but maintains a lower ratio of atmospheric to planetary surface absorption. The reduction in albedo leads to planets orbiting cooler stars to be globally warmer, with more efficient atmospheric transport of heat and moisture from the dayside to the nightside, due to stronger equatorial jets. This decreases the ratio of dayside to nightside OLR. We find that the combined contribution of water vapour and cloud to the nightside radiator fin effect is enhanced to a similar degree for all stellar types, contributing \textasciitilde$20\%$ of the nightside OLR for M-- and G--dwarfs. Overall, for planets near the outer edge of the habitable zone, and with an Earth--like composition, those orbiting cooler stars may be considered more habitable than similar counterparts orbiting hotter stars, as they are likely to have a larger surface region that can support liquid water.
    
    It is important to note that all our simulations adopt the current planetary parameters estimated for TRAPPIST--1e, with the stellar spectrum then varied. The total stellar irradiance is held constant by varying the orbital semi--major axis only between simulations, yet a tidally--locked configuration is retained. Therefore, the two additional simulations irradiated by Proxima Centauri and the Sun are not designed to represent any real planet, and indeed the resulting rotation rate will be inconsistent with the orbital period and tidally--locked state (as the rotation period should increase with semi--major axis for a tidally--locked planet). We have designed our simulations such as to isolate the impacts of a different host star spectrum on the simulated planetary climates.
 
    Our methodology, however, has important limitations which must be addressed with additional research beyond the scope of this study. As M--dwarfs are more active than G--dwarfs, and the planet must orbit closer to the host star to intercept similar stellar flux levels, the impact of flares and high energy radiation must be considered \citep[see for example][]{tilley_2019}. Concerted studies in 3D are required to explore the interaction of the stellar activity with the atmosphere, and in particular the potential impacts on the O$_3$ distribution, given that tidal locking gives rise to a permanent day and night side, the latter never receiving direct stellar radiation. This work has begun for quiescent host stars \citep{yates20}, and we are working on extending it to include host star activity. Furthermore, in our simulations the atmospheric composition has been kept constant in a simplified Earth--like configuration. It is clear from our own Solar System that terrestrial planets can have extremely different compositions. As in this study we are focused on the impacts of stellar spectra on climate and potential implications on habitability, we base our work on the only, currently known inhabited planet, Earth. However, Earth has sustained life through very different atmospheric compositions given the the first evidence of life on Earth is from at least as early as 3.7 Ga \citep{rosing99,hassenkam17}.
 
    Additionally, potentially important climate processes/mechanisms have also been omitted such as atmospheric chemistry \citep[e.g. ozone in][]{yates20}, land--surface impact \citep[e.g.][]{lewis18}, dust \citep[e.g.][]{boutle20}, ocean heat transport \citep[e.g.][]{yang14,yang19b,delgenio19} and, perhaps sea or land ice \citep{rose17}. In particular, \citet{yang14} and \citet{delgenio19} found that ocean transport also acts to reduce the day--night temperature contrast. Inclusion of ocean heat transport would be expected to decrease our predicted day--night temperature contrasts and dayside convection. However, the ocean transport is sensitive to the land/ocean configuration \citep{yang19b}. Additionally, ice formation may lead to a cooling of all our simulations, if it were included, and potentially increase the differences between the M--dwarf cases and the simulation using the Sun, due to the higher ice albedo under G--dwarf stellar spectra \citep{shields13}. However, as ice formation might well be limited to the nightside, its effect on the overall climate of a tidally--locked planet could be small. The reduction in ice albedo when moving from hotter to cooler host stars suggests their orbiting planets may be more resistant to entering a ``snowball'' state \citep{rushby19}, which has occurred at least three times for Earth \citep[e.g.][]{lenton11}. Several studies have questioned whether the climate of tidally--locked planets can exist in a stable regime and avoid atmospheric collapse \citep{Kasting14, turbet18}. The reduced day--night temperature contrast found in our simulations irradiated by cooler host stars may aid their atmospheric stability.
  
    The adoption of a fixed, Earth--like atmospheric composition also neglects the impact of the stellar irradiation on the long--term evolution of the atmosphere, which is required to determine the likely atmospheric composition. However, this is a difficult and poorly constrained problem \citep{bolmont17,dong18}. Finally, we have not considered the impact of life itself. The presence of life on terrestrial exoplanets may fundamentally alter the atmospheric composition \citep{nicholson18,vecchio20}, as has likely happened throughout Earth's own history \citep{lenton11, lenton18}. We must also consider that photosynthesis on Earth is highly adapted towards the spectra it receives and the consequences this may have on the evolution of life \citep{lingam19,lingam20}.

\begin{acknowledgements}
    We are grateful to Professor Jun Yang, whose comments during the review process significantly improved this manuscript. JE would like to thank the Hill Family Scholarship. The Hill Family Scholarship has been generously supported by University of Exeter alumnus, and president of the University’s US Foundation Graham Hill (Economic \& Political Development, 1992) and other donors to the US Foundation. Material produced using Met Office Software. NM and TL gratefully acknowledge funding from a Leverhulme Trust Research Project Grant. JM and IB acknowledge the support of a Met Office Academic Partnership secondment. We acknowledge use of the Monsoon system, a collaborative facility supplied under the Joint Weather and Climate Research Programme, a strategic partnership between the Met Office and the Natural Environment Research Council. This research made use of the ISCA High Performance Computing Service at the University of Exeter. This work was performed using the DiRAC Data Intensive service at Leicester, operated by the University of Leicester IT Services, which forms part of the STFC DiRAC HPC Facility (www.dirac.ac.uk). The equipment was funded by BEIS capital funding via STFC capital grants ST/K000373/1 and ST/R002363/1 and STFC DiRAC Operations grant ST/R001014/1. DiRAC is part of the National e-Infrastructure. This work was partly supported by a Science and Technology Facilities Council Consolidated Grant (ST/R000395/1). We would like to thank David Admundsen and NASA GISS for the use of their spectral files in these simulations.
\end{acknowledgements}

%
\bibliographystyle{aa} 
\bibliography{stellar_comp} 

\begin{thebibliography}{85}
\expandafter\ifx\csname natexlab\endcsname\relax\def\natexlab#1{#1}\fi

\bibitem[{{Amundsen} {et~al.}(2014){Amundsen}, {Baraffe}, {Tremblin},
  {Manners}, {Hayek}, {Mayne}, \& {Acreman}}]{amundsen14}
{Amundsen}, D.~S., {Baraffe}, I., {Tremblin}, P., {et~al.} 2014, \aap, 564, A59

\bibitem[{{Amundsen} {et~al.}(2016){Amundsen}, {Mayne}, {Baraffe}, {Manners},
  {Tremblin}, {Drummond}, {Smith}, {Acreman}, \& {Homeier}}]{amundsen16}
{Amundsen}, D.~S., {Mayne}, N.~J., {Baraffe}, I., {et~al.} 2016, \aap, 595, A36

\bibitem[{{Amundsen} {et~al.}(2017){Amundsen}, {Tremblin}, {Manners},
  {Baraffe}, \& {Mayne}}]{amundsen17}
{Amundsen}, D.~S., {Tremblin}, P., {Manners}, J., {Baraffe}, I., \& {Mayne},
  N.~J. 2017, \aap, 598, A97

\bibitem[{Anglada-Escud{\'e} {et~al.}(2016)Anglada-Escud{\'e}, Amado, Barnes,
  Berdi{\~n}as, Butler, Coleman, de~la Cueva, Dreizler, Endl, Giesers, Jeffers,
  Jenkins, Jones, Kiraga, K{\"u}rster, L{\'o}pez-Gonz{\'a}lez, Marvin, Morales,
  Morin, Nelson, Ortiz, Ofir, Paardekooper, Reiners, Rodr{\'\i}guez,
  Rodrίguez-L{\'o}pez, Sarmiento, Strachan, Tsapras, Tuomi, \&
  Zechmeister}]{anglada16}
Anglada-Escud{\'e}, G., Amado, P.~J., Barnes, J., {et~al.} 2016, Nature, 536,
  437

\bibitem[{{Bolmont} {et~al.}(2017){Bolmont}, {Selsis}, {Owen}, {Ribas},
  {Raymond}, {Leconte}, \& {Gillon}}]{bolmont17}
{Bolmont}, E., {Selsis}, F., {Owen}, J.~E., {et~al.} 2017, \mnras, 464, 3728

\bibitem[{{Boutle} {et~al.}(2020){Boutle}, {Joshi}, {Lambert}, {Lyster},
  {Manners}, {Mayne}, {Ridgway}, {et~al.}}]{boutle20}
{Boutle}, I.~A., {Joshi}, M., {Lambert}, F.~H., {et~al.} 2020, accepted for
  publication in Nature Communications

\bibitem[{{Boutle} {et~al.}(2017){Boutle}, {Mayne}, {Drummond}, {Manners},
  {Goyal}, {Hugo Lambert}, {Acreman}, \& {Earnshaw}}]{boutle17}
{Boutle}, I.~A., {Mayne}, N.~J., {Drummond}, B., {et~al.} 2017, \aap, 601, A120

\bibitem[{Brown {et~al.}(2008)Brown, Beare, Edwards, Lock, Keogh, Milton, \&
  Walters}]{brown08}
Brown, A.~R., Beare, R.~J., Edwards, J.~M., {et~al.} 2008, Boundary-Layer
  Meteorology, 128, 117

\bibitem[{{Debras} {et~al.}(2019){Debras}, {Mayne}, {Baraffe}, {Goffrey}, \&
  {Thuburn}}]{debras19}
{Debras}, F., {Mayne}, N., {Baraffe}, I., {Goffrey}, T., \& {Thuburn}, J. 2019,
  \aap, 631, A36

\bibitem[{{Debras} {et~al.}(2020){Debras}, {Mayne}, {Baraffe}, {Jaupart},
  {Mourier}, {Laibe}, {Goffrey}, \& {Thuburn}}]{debras20}
{Debras}, F., {Mayne}, N., {Baraffe}, I., {et~al.} 2020, \aap, 633, A2

\bibitem[{{Del Genio} {et~al.}(2019){Del Genio}, {Way}, {Amundsen}, {Aleinov},
  {Kelley}, {Kiang}, \& {Clune}}]{delgenio19}
{Del Genio}, A.~D., {Way}, M.~J., {Amundsen}, D.~S., {et~al.} 2019,
  Astrobiology, 19, 99, pMID: 30183335

\bibitem[{Dong {et~al.}(2018)Dong, Jin, Lingam, Airapetian, Ma, \& van~der
  Holst}]{dong18}
Dong, C., Jin, M., Lingam, M., {et~al.} 2018, Proceedings of the National
  Academy of Sciences, 115, 260

\bibitem[{{Donohoe} \& {Battisti}(2011)}]{donohoe11}
{Donohoe}, A. \& {Battisti}, D.~S. 2011, Journal of Climate, 24, 4402

\bibitem[{{Drummond} {et~al.}(2018{\natexlab{a}}){Drummond}, {Mayne},
  {Baraffe}, {Tremblin}, {Manners}, {Amundsen}, {Goyal}, \&
  {Acreman}}]{drummond18b}
{Drummond}, B., {Mayne}, N.~J., {Baraffe}, I., {et~al.} 2018{\natexlab{a}},
  \aap, 612, A105

\bibitem[{{Drummond} {et~al.}(2018{\natexlab{b}}){Drummond}, {Mayne},
  {Manners}, {Baraffe}, {Goyal}, {Tremblin}, {Sing}, \& {Kohary}}]{drummond18c}
{Drummond}, B., {Mayne}, N.~J., {Manners}, J., {et~al.} 2018{\natexlab{b}},
  \apj, 869, 28

\bibitem[{{Drummond} {et~al.}(2018{\natexlab{c}}){Drummond}, {Mayne},
  {Manners}, {Carter}, {Boutle}, {Baraffe}, {H{\'e}brard}, {Tremblin}, {Sing},
  {Amundsen}, \& {Acreman}}]{drummond18a}
{Drummond}, B., {Mayne}, N.~J., {Manners}, J., {et~al.} 2018{\natexlab{c}},
  \apjl, 855, L31

\bibitem[{{Drummond, Benjamin} {et~al.}(2020){Drummond, Benjamin}, {H\'ebrard,
  Eric}, {Mayne, Nathan J.}, {Venot, Olivia}, {Ridgway, Robert J.}, {Changeat,
  Quentin}, {Tsai, Shang-Min}, {Manners, James}, {Tremblin, Pascal}, {Abraham,
  Nathan Luke}, {Sing, David}, \& {Kohary, Krisztian}}]{drummond20}
{Drummond, Benjamin}, {H\'ebrard, Eric}, {Mayne, Nathan J.}, {et~al.} 2020,
  \aap, 636, A68

\bibitem[{Fauchez {et~al.}(2020)Fauchez, Turbet, Wolf, Boutle, Way, Del~Genio,
  Mayne, Tsigaridis, Kopparapu, Yang, Forget, Mandell, \&
  Domagal~Goldman}]{fauchez20}
Fauchez, T., Turbet, M., Wolf, E.~T., {et~al.} 2020, Geosci. Model Dev., 13,
  707

\bibitem[{Frierson {et~al.}(2006)Frierson, Held, \& Zurita-Gotor}]{frierson06}
Frierson, D. M.~W., Held, I.~M., \& Zurita-Gotor, P. 2006, Journal of the
  Atmospheric Sciences, 63, 2548

\bibitem[{Gillon {et~al.}(2017)Gillon, Triaud, Demory, Jehin, Agol, Deck,
  Lederer, de~Wit, Burdanov, Ingalls, Bolmont, Leconte, Raymond, Selsis,
  Turbet, Barkaoui, Burgasser, Burleigh, Carey, Chaushev, Copperwheat, Delrez,
  Fernandes, Holdsworth, Kotze, Van~Grootel, Almleaky, Benkhaldoun, Magain, \&
  Queloz}]{gillon17}
Gillon, M., Triaud, A. H. M.~J., Demory, B., {et~al.} 2017, Nature, 542, 456

\bibitem[{Gregory \& Rowntree(1990)}]{gregory90}
Gregory, D. \& Rowntree, P.~R. 1990, Monthly Weather Review, 118, 1483

\bibitem[{Grimm {et~al.}(2018)Grimm, Demory, Gillon, Dorn, Agol, Burdanov,
  Delrez, Sestovic, Triaud, {Turbet, M.}, {Bolmont, \'E.}, {Caldas, A.}, {de
  Wit, J.}, {Jehin, E.}, {Leconte, J.}, {Raymond, S. N.}, {Van Grootel, V.},
  {Burgasser, A. J.}, {Carey, S.}, {Fabrycky, D.}, {Heng, K.}, {Hernandez, D.
  M.}, {Ingalls, J. G.}, {Lederer, S.}, {Selsis, F.}, \& {Queloz,
  D.}}]{grimm18}
Grimm, S.~L., Demory, B., Gillon, M., {et~al.} 2018, A\&A, 613, A68

\bibitem[{Haqq-Misra {et~al.}(2018)Haqq-Misra, Wolf, Joshi, Zhang, \&
  Kopparapu}]{haqq-misra2018}
Haqq-Misra, J., Wolf, E.~T., Joshi, M., Zhang, X., \& Kopparapu, R.~K. 2018,
  The Astrophysical Journal, 852, 67

\bibitem[{Hassenkam {et~al.}(2017)Hassenkam, Andersson, Dalby, Mackenzie, \&
  Rosing}]{hassenkam17}
Hassenkam, T., Andersson, M.~P., Dalby, K.~N., Mackenzie, D. M.~A., \& Rosing,
  M.~T. 2017, Nature, 548, 78

\bibitem[{{Helling} {et~al.}(2016){Helling}, {Lee}, {Dobbs-Dixon}, {Mayne},
  {Amundsen}, {Khaimova}, {Unger}, {Manners}, {Acreman}, \&
  {Smith}}]{helling16}
{Helling}, C., {Lee}, G., {Dobbs-Dixon}, I., {et~al.} 2016, \mnras, 460, 855

\bibitem[{{Howard} {et~al.}(2018){Howard}, {Tilley}, {Corbett}, {Youngblood},
  {Loyd}, {Ratzloff}, {Law}, {Fors}, {del Ser}, {Shkolnik}, {Ziegler}, {Goeke},
  {Pietraallo}, \& {Haislip}}]{howard_2018}
{Howard}, W.~S., {Tilley}, M.~A., {Corbett}, H., {et~al.} 2018, \apjl, 860, L30

\bibitem[{Hunter(2007)}]{matplotlib}
Hunter, J.~D. 2007, Computing in Science \& Engineering, 9, 90

\bibitem[{Jin {et~al.}(2011)Jin, Qiao, Wang, Fang, \& Yi}]{jin11}
Jin, Z., Qiao, Y., Wang, Y., Fang, Y., \& Yi, W. 2011, Opt. Express, 19, 26429

\bibitem[{Joshi {et~al.}(2020)Joshi, Elvidge, Wordsworth, \& Sergeev}]{joshi20}
Joshi, M.~M., Elvidge, A.~D., Wordsworth, R., \& Sergeev, D. 2020, \apj, 892,
  L33

\bibitem[{Joshi \& Haberle(2012)}]{joshi12}
Joshi, M.~M. \& Haberle, R.~M. 2012, Astrobiology, 12, 3, pMID: 22181553

\bibitem[{Kasting {et~al.}(2014)Kasting, Kopparapu, Ramirez, \&
  Harman}]{Kasting14}
Kasting, J.~F., Kopparapu, R., Ramirez, R.~M., \& Harman, C.~E. 2014,
  Proceedings of the National Academy of Sciences, 111, 12641

\bibitem[{{Kasting} {et~al.}(1993){Kasting}, {Whitmire}, \&
  {Reynolds}}]{kasting93}
{Kasting}, J.~F., {Whitmire}, D.~P., \& {Reynolds}, R.~T. 1993, \icarus, 101,
  108

\bibitem[{Koll \& Abbot(2016)}]{koll16}
Koll, D. D.~B. \& Abbot, D.~S. 2016, \apj, 825, 99

\bibitem[{Komacek \& Abbot(2019)}]{komacek19}
Komacek, T.~D. \& Abbot, D.~S. 2019, \apj, 871, 245

\bibitem[{Lenton {et~al.}(2018)Lenton, Daines, \& Mills}]{lenton18}
Lenton, T.~M., Daines, S.~J., \& Mills, B.~J. 2018, Earth-Science Reviews, 178,
  1

\bibitem[{Lenton \& Watson(2011)}]{lenton11}
Lenton, T.~M. \& Watson, A.~J. 2011, Revolutions that made the Earth (Oxford
  University Press)

\bibitem[{{Lewis} {et~al.}(2018){Lewis}, {Lambert}, {Boutle}, {Mayne},
  {Manners}, \& {Acreman}}]{lewis18}
{Lewis}, N.~T., {Lambert}, F.~H., {Boutle}, I.~A., {et~al.} 2018, \apj, 854,
  171

\bibitem[{{Lines} {et~al.}(2018{\natexlab{a}}){Lines}, {Manners}, {Mayne},
  {Goyal}, {Carter}, {Boutle}, {Lee}, {Helling}, {Drummond}, {Acreman}, \&
  {Sing}}]{lines18b}
{Lines}, S., {Manners}, J., {Mayne}, N.~J., {et~al.} 2018{\natexlab{a}},
  \mnras, 481, 194

\bibitem[{{Lines} {et~al.}(2018{\natexlab{b}}){Lines}, {Mayne}, {Boutle},
  {Manners}, {Lee}, {Helling}, {Drummond}, {Amundsen}, {Goyal}, {Acreman},
  {Tremblin}, \& {Kerslake}}]{lines18a}
{Lines}, S., {Mayne}, N.~J., {Boutle}, I.~A., {et~al.} 2018{\natexlab{b}},
  \aap, 615, A97

\bibitem[{Lines {et~al.}(2019)Lines, Mayne, Manners, Boutle, Drummond,
  Mikal-Evans, Kohary, \& Sing}]{lines19}
Lines, S., Mayne, N.~J., Manners, J., {et~al.} 2019, \mnras, 488, 1332

\bibitem[{Lingam \& Loeb(2019)}]{lingam19}
Lingam, M. \& Loeb, A. 2019, \mnras, 485, 5924–5928

\bibitem[{{Lingam} \& {Loeb}(2020)}]{lingam20}
{Lingam}, M. \& {Loeb}, A. 2020, \apjl, 889, L15

\bibitem[{Lock {et~al.}(2000)Lock, Brown, Bush, Martin, \& Smith}]{lock00}
Lock, A.~P., Brown, A.~R., Bush, M.~R., Martin, G.~M., \& Smith, R. N.~B. 2000,
  Monthly Weather Review, 128

\bibitem[{{Mayne} {et~al.}(2014{\natexlab{a}}){Mayne}, {Baraffe}, {Acreman},
  {Smith}, {Browning}, {Sk{\aa}lid Amundsen}, {Wood}, {Thuburn}, \&
  {Jackson}}]{mayne14a}
{Mayne}, N.~J., {Baraffe}, I., {Acreman}, D.~M., {et~al.} 2014{\natexlab{a}},
  \aap, 561, A1

\bibitem[{{Mayne} {et~al.}(2014{\natexlab{b}}){Mayne}, {Baraffe}, {Acreman},
  {Smith}, {Wood}, {Amundsen}, {Thuburn}, \& {Jackson}}]{mayne14b}
{Mayne}, N.~J., {Baraffe}, I., {Acreman}, D.~M., {et~al.} 2014{\natexlab{b}},
  Geoscientific Model Development, 7, 3059

\bibitem[{{Mayne} {et~al.}(2017){Mayne}, {Debras}, {Baraffe}, {Thuburn},
  {Amundsen}, {Acreman}, {Smith}, {Browning}, {Manners}, \& {Wood}}]{mayne17}
{Mayne}, N.~J., {Debras}, F., {Baraffe}, I., {et~al.} 2017, \aap, 604, A79

\bibitem[{{Mayne} {et~al.}(2019){Mayne}, {Drummond}, {Debras}, {Jaupart},
  {Manners}, {Boutle}, {Baraffe}, \& {Kohary}}]{mayne19}
{Mayne}, N.~J., {Drummond}, B., {Debras}, F., {et~al.} 2019, \apj, 871, 56

\bibitem[{Merlis \& Schneider(2010)}]{merlis10}
Merlis, T.~M. \& Schneider, T. 2010, Journal of Advances in Modeling Earth
  Systems, 2

\bibitem[{{Met Office}(2010-2020)}]{iris}
{Met Office}. 2010-2020, {Iris: A Python library for analysing and visualising
  meteorological and oceanographic data sets}

\bibitem[{{Nicholson} {et~al.}(2018){Nicholson}, {Wilkinson}, Hywel T
  P~{Williams}, \& {Lenton}}]{nicholson18}
{Nicholson}, A.~E., {Wilkinson}, D.~M., Hywel T P~{Williams}, H.~T.~P., \&
  {Lenton}, T.~M. 2018, Monthly Notices of the Royal Astronomical Society, 477,
  727

\bibitem[{Penn \& Vallis(2018)}]{penn18}
Penn, J. \& Vallis, G.~K. 2018, \apj, 868, 147

\bibitem[{Pierrehumbert(2010)}]{pierrehumbert10}
Pierrehumbert, R.~T. 2010, Principles of {P}lanetary {C}limate (Cambridge
  University Press)

\bibitem[{Pierrehumbert \& Hammond(2019)}]{pierrehumbert19}
Pierrehumbert, R.~T. \& Hammond, M. 2019, Annual Review of Fluid Mechanics, 51,
  275

\bibitem[{Polyansky {et~al.}(2018)Polyansky, Kyuberis, Zobov, Tennyson,
  Yurchenko, \& Lodi}]{polyansky18}
Polyansky, O.~L., Kyuberis, A.~A., Zobov, N.~F., {et~al.} 2018, Monthly Notices
  of the Royal Astronomical Society, 480, 2597

\bibitem[{{Rajpurohit} {et~al.}(2013){Rajpurohit}, {Reyl{\'e}}, {Allard},
  {Homeier}, {Schultheis}, {Bessell}, \& {Robin}}]{rajpurohit13}
{Rajpurohit}, A.~S., {Reyl{\'e}}, C., {Allard}, F., {et~al.} 2013, \aap, 556

\bibitem[{Rose {et~al.}(2017)Rose, Cronin, \& Bitz}]{rose17}
Rose, B. E.~J., Cronin, T.~W., \& Bitz, C.~M. 2017, The Astrophysical Journal,
  846, 28

\bibitem[{Rosing(1999)}]{rosing99}
Rosing, M.~T. 1999, Science, 283, 674

\bibitem[{Rushby {et~al.}(2019)Rushby, Shields, \& Joshi}]{rushby19}
Rushby, A.~J., Shields, A.~L., \& Joshi, M. 2019, \apj, 887, 29

\bibitem[{{Sainsbury-Martinez} {et~al.}(2019){Sainsbury-Martinez}, {Wang},
  {Fromang}, {Tremblin}, {Dubos}, {Meurdesoif}, {Spiga}, {Leconte}, {Baraffe},
  {Chabrier}, {Mayne}, {Drummond}, \& {Debras}}]{sainsbury19}
{Sainsbury-Martinez}, F., {Wang}, P., {Fromang}, S., {et~al.} 2019, \aap, 632,
  A114

\bibitem[{Schlaufman \& Laughlin(2010)}]{schlaufman10}
Schlaufman, K.~C. \& Laughlin, G. 2010, \aap, 519

\bibitem[{Sergeev {et~al.}(2020)Sergeev, Lambert, Mayne, Boutle, Manners, \&
  Kohary}]{sergeev20}
Sergeev, D.~E., Lambert, F.~H., Mayne, N.~J., {et~al.} 2020, \apj, 894, 84

\bibitem[{Shields {et~al.}(2019)Shields, Bitz, \& Palubski}]{shields19}
Shields, A.~L., Bitz, C.~M., \& Palubski, I. 2019, \apj, 884, L2

\bibitem[{Shields {et~al.}(2013)Shields, Meadows, Bitz, Pierrehumbert, Joshi,
  \& Robinson}]{shields13}
Shields, A.~L., Meadows, V.~S., Bitz, C.~M., {et~al.} 2013, Astrobiology, 13,
  715, pMID: 23855332

\bibitem[{{Showman} \& {Polvani}(2010)}]{showman10}
{Showman}, A.~P. \& {Polvani}, L.~M. 2010, \grl, 37, L18811

\bibitem[{{Showman} \& {Polvani}(2011)}]{showman11}
{Showman}, A.~P. \& {Polvani}, L.~M. 2011, \apj, 738, 71

\bibitem[{Tashkun \& Perevalov(2011)}]{tashkun11}
Tashkun, S. \& Perevalov, V. 2011, Journal of Quantitative Spectroscopy and
  Radiative Transfer, 112, 1403

\bibitem[{Tennyson {et~al.}(2016)Tennyson, Yurchenko, Al-Refaie, Barton, Chubb,
  Coles, Diamantopoulou, Gorman, Hill, Lam, Lodi, McKemmish, Na, Owens,
  Polyansky, Rivlin, Sousa-Silva, Underwood, Yachmenev, \& Zak}]{tennyson16}
Tennyson, J., Yurchenko, S.~N., Al-Refaie, A.~F., {et~al.} 2016, Journal of
  Molecular Spectroscopy, 327, 73 , new Visions of Spectroscopic Databases,
  Volume II

\bibitem[{Thomson \& Vallis(2019)}]{thomson19}
Thomson, S.~I. \& Vallis, G.~K. 2019, Quarterly Journal of the Royal
  Meteorological Society, 145, 2627

\bibitem[{Tilley {et~al.}(2019)Tilley, Segura, Meadows, Hawley, \&
  Davenport}]{tilley_2019}
Tilley, M.~A., Segura, A., Meadows, V., Hawley, S., \& Davenport, J. 2019,
  Astrobiology, 19, 64, pMID: 30070900

\bibitem[{{Tremblin} {et~al.}(2017){Tremblin}, {Chabrier}, {Mayne}, {Amundsen},
  {Baraffe}, {Debras}, {Drummond}, {Manners}, \& {Fromang}}]{tremblin17}
{Tremblin}, P., {Chabrier}, G., {Mayne}, N.~J., {et~al.} 2017, \apj, 841, 30

\bibitem[{{Turbet} {et~al.}(2018){Turbet}, {Bolmont}, {Leconte}, {Forget},
  {Selsis}, {Tobie}, {Caldas}, {Naar}, \& {Gillon}}]{turbet18}
{Turbet}, M., {Bolmont}, E., {Leconte}, J., {et~al.} 2018, \aap, 612, A86

\bibitem[{{Turbet} {et~al.}(2016){Turbet}, {Leconte}, {Selsis}, {Bolmont},
  {Forget}, {Ribas}, {Raymond}, \& {Anglada-Escud\'e}}]{turbet16}
{Turbet}, M., {Leconte}, J., {Selsis}, F., {et~al.} 2016, \aap, 596, A112

\bibitem[{Vecchio {et~al.}(2020)Vecchio, Primavera, Lepreti, Alberti, \&
  Carbone}]{vecchio20}
Vecchio, A., Primavera, L., Lepreti, F., Alberti, T., \& Carbone, V. 2020, The
  Astrophysical Journal, 891, 24

\bibitem[{Walters {et~al.}(2019)Walters, Baran, Boutle, Brooks, Earnshaw,
  Edwards, Furtado, Hill, Lock, Manners, Morcrette, Mulcahy, Sanchez, Smith,
  Stratton, Tennant, Tomassini, Van~Weverberg, Vosper, Willett, Browse,
  Bushell, Carslaw, Dalvi, Essery, Gedney, Hardiman, Johnson, Johnson, Jones,
  Jones, Mann, Milton, Rumbold, Sellar, Ujiie, Whitall, Williams, \&
  Zerroukat}]{walters19}
Walters, D., Baran, A.~J., Boutle, I., {et~al.} 2019, Geoscientific Model
  Development, 12, 1909

\bibitem[{Wilson \& Ballard(1999)}]{wilson99}
Wilson, D.~R. \& Ballard, S.~P. 1999, Quarterly Journal of the Royal
  Meteorological Society, 125, 1607

\bibitem[{Wilson {et~al.}(2008)Wilson, Bushell, Kerr-Munslow, Price, \&
  Morcrette}]{wilson08}
Wilson, D.~R., Bushell, A.~C., Kerr-Munslow, A.~M., Price, J.~D., \& Morcrette,
  C.~J. 2008, Quarterly Journal of the Royal Meteorological Society, 134, 2093

\bibitem[{Wolf(2017)}]{wolf17b}
Wolf, E.~T. 2017, The Astrophysical Journal, 839, L1

\bibitem[{Wood {et~al.}(2014)Wood, Staniforth, White, Allen, Diamantakis,
  Gross, Melvin, Smith, Vosper, Zerroukat, \& Thuburn}]{wood14}
Wood, N., Staniforth, A., White, A., {et~al.} 2014, Quarterly Journal of the
  Royal Meteorological Society, 140, 1505

\bibitem[{{Yang} \& {Yang}(2019)}]{yang19}
{Yang}, H. \& {Yang}, J. 2019, arXiv e-prints, arXiv:1910.06479

\bibitem[{Yang \& Abbot(2014)}]{yang14}
Yang, J. \& Abbot, D.~S. 2014, \apj, 784, 155

\bibitem[{Yang {et~al.}(2019{\natexlab{a}})Yang, Abbot, Koll, Hu, \&
  Showman}]{yang19b}
Yang, J., Abbot, D.~S., Koll, D. D.~B., Hu, Y., \& Showman, A.~P.
  2019{\natexlab{a}}, \apj, 871, 29

\bibitem[{Yang {et~al.}(2013)Yang, Cowan, \& Abbot}]{yang13}
Yang, J., Cowan, N.~B., \& Abbot, D.~S. 2013, \apjl, 771, L45

\bibitem[{Yang {et~al.}(2019{\natexlab{b}})Yang, Leconte, Wolf, Merlis, Koll,
  Forget, \& Abbot}]{yang19c}
Yang, J., Leconte, J., Wolf, E.~T., {et~al.} 2019{\natexlab{b}}, The
  Astrophysical Journal, 875, 46

\bibitem[{{Yates} {et~al.}(2020){Yates}, {Palmer}, {Manners}, {Boutle},
  {Kohary}, {Mayne}, \& {Abraham}}]{yates20}
{Yates}, J.~S., {Palmer}, P.~I., {Manners}, J., {et~al.} 2020, \mnras, 492,
  1691

\bibitem[{Yurchenko {et~al.}(2018)Yurchenko, Al-Refaie, \&
  Tennyson}]{Yurchenko18}
Yurchenko, S.~N., Al-Refaie, A.~F., \& Tennyson, J. 2018, Astronomy \&
  Astrophysics, 614, A131

\end{thebibliography}
%

\end{document}